\pdfoutput=1
\documentclass[pre,superscriptaddress,twocolumn]{revtex4}
\usepackage{latexsym}
\usepackage{graphicx,amsmath,color}

\newcommand{\rmd}{\mathrm{d}}
\newcommand{\rme}{\mathrm{e}}

\renewcommand{\topfraction}{1.0}

\newlength{\figsize}
\begin{document}

\bibliographystyle{KAY}

\title{\sffamily \bfseries \Large Avalanche-size distribution at the
depinning transition: \\
A numerical test of the theory}
\author{\sffamily \bfseries Alberto Rosso}
 \affiliation{LPTMS; CNRS and Universite Paris-Sud, UMR 8626, 91405 Orsay Cedex, France.} \affiliation{CNRS-Laboratoire de Physique
Th{\'e}orique de l'Ecole Normale Sup{\'e}rieure, 24 rue Lhomond, 75231
Paris Cedex, France.} 
\author{\sffamily \bfseries Pierre Le Doussal} \affiliation{CNRS-Laboratoire de Physique
Th{\'e}orique de l'Ecole Normale Sup{\'e}rieure, 24 rue Lhomond, 75231
Paris Cedex, France.} 
\author{\sffamily \bfseries Kay
J\"org Wiese} \affiliation{CNRS-Laboratoire de Physique
Th{\'e}orique de l'Ecole Normale Sup{\'e}rieure, 24 rue Lhomond, 75231
Paris Cedex, France.}
\smallskip

\date{\small\today}
\begin{abstract}
We calculate numerically the sizes $S$ of jumps (avalanches) between successively pinned configurations of
an elastic line ($d=1$) or interface ($d=2$), pulled by a spring of (small) strength $m^2$ in a random-field landscape. 
We 
obtain strong evidence that the size distribution, away from the small-scale cutoff, takes the form $P(S)  = \frac{\left< S \right>}{S_m^2} p(S/S_{m})$ where $S_m:=\frac{\left <S^2 \right>}{2 \left<S\right>}\sim m^{-d-\zeta}$ is the scale of avalanches, and $\zeta$ the roughness exponent at the depinning transition. Measurement of the scaling function $f(s) := s^{\tau} p(s) $ is compared with the predictions from a recent Functional RG (FRG) calculation, both at mean-field and one-loop level. The avalanche-size exponent $\tau$ is found in good agreement with the conjecture $\tau =2- 2/(d+\zeta)$, recently confirmed to one loop via the FRG. The function $f(s)$ exhibits a shoulder and a stretched exponential decay at large $s$, $\ln f(s) \sim - s^{\delta}$, with $\delta \approx 7/6$ in $d=1$. The function $f(s)$,  universal ratios of moments, and the generating function $\left<e^{\lambda s}\right>$ are found in excellent agreement with the one-loop FRG predictions. The distribution of {\em local} avalanche sizes $S_\phi$, i.e.\ of the jumps of a subspace of the manifold of dimension $d_\phi$, is also computed and compared to our FRG predictions, and to the conjecture $\tau_\phi =2- 2/(d_\phi+\zeta)$. 
\end{abstract}
\maketitle

\section{Introduction}\label{a1}

Elastic objects pinned by a random substrate are ubiquitous in nature. The competition between elastic restoring forces and quenched disorder results in multiple metastable states. Upon applying an external force one observes collective jerky motion which proceeds by sudden jumps, called avalanches. Examples are the Barkhausen noise in magnets
\cite{UrbachMadisonMarkert1995,MehtaMillsDahmenSethna2002,DahmenSethnaKuntzPerkovic2001,CarpenterDahmenSethnaFriedmanLoverdeVanderveld2001,DahmenSethnaPerkovic2000,PerkovicDahmenSethna1995,DahmenKarthaKrumhanslRobertsSethnaShore1994},  jumps in the creep motion of
magnetic domain walls \cite{LemerleFerreChappertMathetGiamarchiLeDoussal1998,RepainBauerJametFerreMouginChappertBernas2004,MetaxasJametMouginCormierFerreBaltzRodmacqDienyStamps2007,SethnaDahmenMyers2001}, avalanches in the depinning of a 
contact-line of a fluid
\cite{PrevostRolleyGuthmann2002,PrevostRolleyGuthmann1999,MoulinetGuthmannRolley2002,MoulinetRossoKrauthRolley2004}, or in dislocation and crack propagation \cite{MorettiMiguelZaiserMoretti2004,PonsonBonamyBouchaud2006,Ponson2007,BonamySantucciPonson2008},
and stick-slip motion of e.g.\ tectonic plates, responsible for earthquakes \cite{FisherDahmenRamanathanBen-Zion1997,DSFisher1998,SchwarzFisher2003,JaglaKolton2009}.
Avalanches have also been studied in models without quenched  substrate disorder, such as 
in sandpile models and in granular matter \cite{TangBak1988,DharRamaswamy1989,BanerjeeSantraBose1995,Dhar2006}.
An important characteristics of avalanche motion is its scale invariance, self-organized criticality, and a broad distribution  $P(S) \sim S^{-\tau}$ of the sizes $S$  of avalanches, for sizes $S$ between a small- and large-scale cutoff $S_{\mathrm{min}} \ll S \ll S_m$. Pinned elastic manifolds are an important prototype of a much wider class of phenomena, reaching far outside physics, e.g.\ into economy and finance, where extreme (and sometimes catastrophic) events are sufficiently frequent and large to dominate most observables. In this context, it is clearly of importance to understand how the avalanche-size probability is cut off at the large scales, for  $S>S_m$. 

Although avalanche motion of pinned manifolds has been studied for a while in numerics \cite{NarayanMiddleton1994,LacombeZapperiHerrmann2001,TanguyGounelleRoux1998}, most work focused on measuring the avalanche-size exponent $\tau$, with minimal guidance from the theory. This is mainly because no analytic approach was available besides mean-field and scaling arguments. The most notable one was proposed by Narayan and Fisher (NF) \cite{NarayanDSFisher1993a} on the basis of the unproved assumption that the avalanche density  remains finite at the depinning threshold, resulting into 
\begin{eqnarray} \label{conj}
\tau=2- \frac{2}{d+\zeta}\ .
\end{eqnarray}
Here $\zeta$ is the roughness exponent at the depinning transition. Progress both in constructing the field theory of the depinning transition \cite{ChauveLeDoussalWiese2000a,LeDoussalWieseChauve2002,LeDoussalWiese2003a,FedorenkoLeDoussalWiese2006}
following the pioneering work on the Functional RG (FRG)\cite{NattermanStepanowTangLeschhorn1992,NarayanDSFisher1993a,LeschhornNattermannStepanow1996}
and in developing new powerful algorithms \cite{RossoKrauth2001b,RossoKrauth2002,RossoHartmannKrauth2002,RossoKrauthLeDoussalVannimenusWiese2003,MoulinetRossoKrauthRolley2004,BolechRosso2004,KoltonRossoGiamarchi2005} 
had focused mostly on structural properties of the pinned manifold, such as the precise determination of $\zeta$. Even an appropriate definition of static and dynamic avalanches, allowing contact with the field theory, had remained unclear. It was given in the statics  \cite{LeDoussal2006b,MiddletonLeDoussalWiese2006,LeDoussal2008} and at depinning \cite{LeDoussalWiese2006a,RossoLeDoussalWiese2006a} using a confining quadratic potential; it led to the measurement, with great accuracy, of the renormalized disorder correlator $\Delta(u)$, i.e.\ the fixed point of the FRG. Only very recently we succeeded in
computing the distribution of avalanche sizes within the FRG \cite{LeDoussalMiddletonWiese2008,LeDoussalWiese2008c,FedorenkoLeDoussalWieseInPrep}. The calculation at tree level gave mean-field predictions (some of them new and non-trivial), valid above the upper critical dimension $d_{\mathrm{uc}}=4$, i.e.\ for $d > 4$.  The one-loop calculation gave an expansion to order $O(\epsilon)$, with $\epsilon=4-d$. Remarkably, the conjecture (\ref{conj}) was confirmed to $O(\epsilon)$ accuracy. It is thus of great interest to test these predictions in numerics. 

The aim of the present paper is to compute numerically the jumps (avalanches) between successively pinned configurations of an elastic line ($d=1$) and interface ($d=2$). The convenient setting to compare with the recent predictions from the FRG is to submit the manifold to an external quadratic well, i.e.\ a spring. We will study mostly  random-field disorder, but we also check that the results are the same for random-bond disorder, as is predicted at depinning and was checked in our previous work \cite{RossoLeDoussalWiese2006a} for the renormalized disorder correlator $\Delta(u)$. Most of the numerical method is similar to our previous work \cite{RossoLeDoussalWiese2006a}.

The outline of this article is as follows: We define  in section \ref{s:2} the model and numerical procedure; and in 
section \ref{s:3} an avalanche, its size, the characteristic scales and the scaling functions. In section \ref{a14} the reader will find our numerical results for the avalanche-size distribution, and their comparison to our analytical results in $d=1$ and $d=2$. In section \ref{s:5} we compare our numerical and analytical results for the universal ratios of algebraic moments, the $r_n$. In section \ref{s:6} we do the same for the generating function of exponential moments, i.e.\ the characteristic function of the size distribution, denoted $\tilde Z(\lambda)$, in $d=1$ and $d=2$. Finally, in Section (\ref{sec:local}), we compute the distribution of local avalanches in $d=1$ and compare with the predictions.

\section{Numerical procedure: parabola and metastable states}\label{s:2}

Let us now describe the model and algorithm, in $d=1$ for simplicity. The procedure is very similar to our previous work
\cite{RossoLeDoussalWiese2006a}. The interface is discretized as $u(x)\equiv u_i$, $i=1,\ldots, L$,  and periodic boundary conditions are taken: $u_0=u_L$, $u_{L+1}=u_1$. We start from a flat interface ($u_i=0$) embedded in a parabolic potential. The
equation of motion is
\begin{equation}
\partial_t u_i=m^2 (w - u_i) + u_{i+1}+u_{i-1}-2 u_i+ F(i,u_i)
\end{equation}
$F(i,u_i)$ is the disorder force.  We distinguish two different
microscopic disorders: 

(i) {\it random force (RF)}: for each integer value of $u_i$ we take
a  random number extracted from a normal distribution. The value of the random
force for non-integer values of $u_i$ is given by the linear
interpolation of the forces at the  two closest integers $u_{i}$. Forces for different $i$ are independent.

(ii) {\it random bond (RB)}: the random force is derived from a random potential:
$F(i,u_i)=-\partial_{u_i} V(i,u_i)$. For each integer value of $u_i$, the potential is a random number normally distributed. The interpolation of $V$
is done by means of a cubic spline connecting $M$ random numbers. Two
extra conditions are needed in order to define a spline: we have taken
$F(i,0)=0$ and $F(i,M)=0$.  In our simulations $M=100$. When the
line advances beyond $u_i=M$, a new spline, with $M$ new random numbers is
generated. Potentials for different $i$ are independent.

The value $w$ is the center of mass of a confining potential for each
point $i$, of the
form 
$\frac{m^{2}}{2} (w-u_{i})^{2}$. In the simulation, $w$ is increased
from $0$. 
For each value of $w$ a metastable state is computed. Increasing $w$,
a stationary sequence of metastable states (independent of the initial
configuration) is reached, as observed in Ref.\ \cite{RossoLeDoussalWiese2006a}.
This is the steady state on which we focus.
Our main results concern an elastic string in $d=1$ of size $L$ with RF disorder, but we have also studied RB disorder, see Fig.  \ref{f:RFconjecture}, and a $2$-dimensional  elastic interface of size $L^2$ with periodic boundary conditions. As expected, for the depinning transition, the RB case falls in the same universality class as the RF case and results are very similar.

\section{Definitions and observables}\label{s:3}

For given $w=w_0$ the manifold
moves to a metastable state $u_{w_0}(x)$, i.e.\ a state dynamically stable to infinitesimally small
deformations. Following the notation of
\cite{LeDoussalMiddletonWiese2008,LeDoussalWiese2008c}, we define the center of mass of the metastable configuration
\begin{equation}
u(w_0):= \frac{1}{L^{d}} \int \rmd x\, u_{w_0}(x) 
\end{equation}
with $L$ the linear size of the system (number of points), and $d$ the
dimension. 
 One then increases $w$, and a  smooth forward
deformation of $u_w(x)$ results (for smooth short-scale disorder)
while the state remains stable. At some $w=w_1$ the state becomes
unstable and the manifold, for $w=w_1^+$ moves until it is blocked
again in a new metastable state $u_{w_1}(x)$ (also locally stable). 
This process is called an avalanche and its size $S$ is defined as the area swept by
the line as it jumps between the two consecutive metastable states: 
 \begin{equation}
S:= L^{d} \left[u(w_1)-u(w_0) \right]
\end{equation}
The distribution of avalanche sizes is expected to exhibit universality, i.e.\ independence of 
short scales, for sizes $S > S_{\mathrm{min}}$. The short-scale cutoff $S_{\mathrm{min}}$  corresponds to  the area spanned by a single monomer on the scale of the discretization of the disorder (in our units $S_{\mathrm{min}}\simeq 1$).
In the limit $m=0$  a critical point is reached, resulting in a power-law distribution of avalanche sizes. To properly define the problem, including the stationary measure, it is essential to consider a small $m>0$. Then, the internal correlation
length $L_m$ is finite: it can e.g. be measured from the structure factor leading to \cite{RossoLeDoussalWiese2006a}
\begin{equation}\label{lm}
L_m \approx 5/m\ .
\end{equation} 
$L_m$ is large in the small-$m$ regime considered here. As a result, the distribution of avalanche sizes is cut off by the large scale $S_m \gg S_{\mathrm{min}}$, defined as
\begin{equation}\label{4}
S_m: =\frac{\langle S^2 \rangle}{ 2 \langle S \rangle}\ .
\end{equation} 
It is expected to scale as $S_m \sim  L_m^{d+\zeta}\sim m^{-d-\zeta}$ at small $m$. Here and below
we define the (normalized) distribution of avalanche sizes $P(S)$, as well as its moments
\begin{equation}\label{a4}
\left< S^{n} \right> := \frac{1}{N}\sum_{i=1}^{N} S_{i}^{n} = \int_0^\infty \rmd S\, S^n P(S) 
\end{equation}
from the sequence of measured avalanches $S_{i}$, $i=1,\dots,N$.

The scale $S_{m}$ is important as it allows to define universal
functions. In the variable $s:=S/S_m$ the avalanche-size distribution
should become universal. Indeed, one of the predictions of the FRG theory is that if the exponent $\tau$ satisfies $2> \tau >1$ which is the
case here, then the distribution of avalanche sizes for $S \gg S_{\mathrm{min}}$ takes the form as $m \to 0$, i.e.\ $S_m \gg S_{\mathrm{min}}$,
\begin{equation}\label{a2}
P (S) \rmd S := \frac{\left< S \right>}{S_m} p \left(\frac{S}{S_{m}}\right)  \frac{\rmd S}{S_m}\ .
\end{equation}
The function $p(s)$ is universal and depends only on the space dimension $d$. Note that the normalized
probability $P(S)$ depends on the cut-off $S_{\mathrm{min}}$ via the
first moment $\left< S \right>$ which cannot be predicted by the theory, hence is
 an input from the numerics. It is important to stress that while the function
$p(s)$ is universal and convenient for data analysis, {\it it is not a probability distribution} and
is not normalized to unity. Rather, it satisfies from its definition (\ref{a2}) and using (\ref{4}) the two normalization conditions
\begin{eqnarray}\label{7}
\left< s \right>_p&=& \int \rmd s\, s p (s) =1\\
\left< s^{2} \right>_p &=&  \int \rmd s\, s^{2} p (s) =2 \ . \label{8} 
\end{eqnarray}
Here and below we use the notation $\left< s \right>_p$ to denote an integration over $p(s)$ and
distinguish it from a true expectation value over $P(S)$, denoted $\left<\ldots\right>$.

\section{The avalanche-size distribution}\label{a14}

\begin{figure}
{\includegraphics[width=\columnwidth]{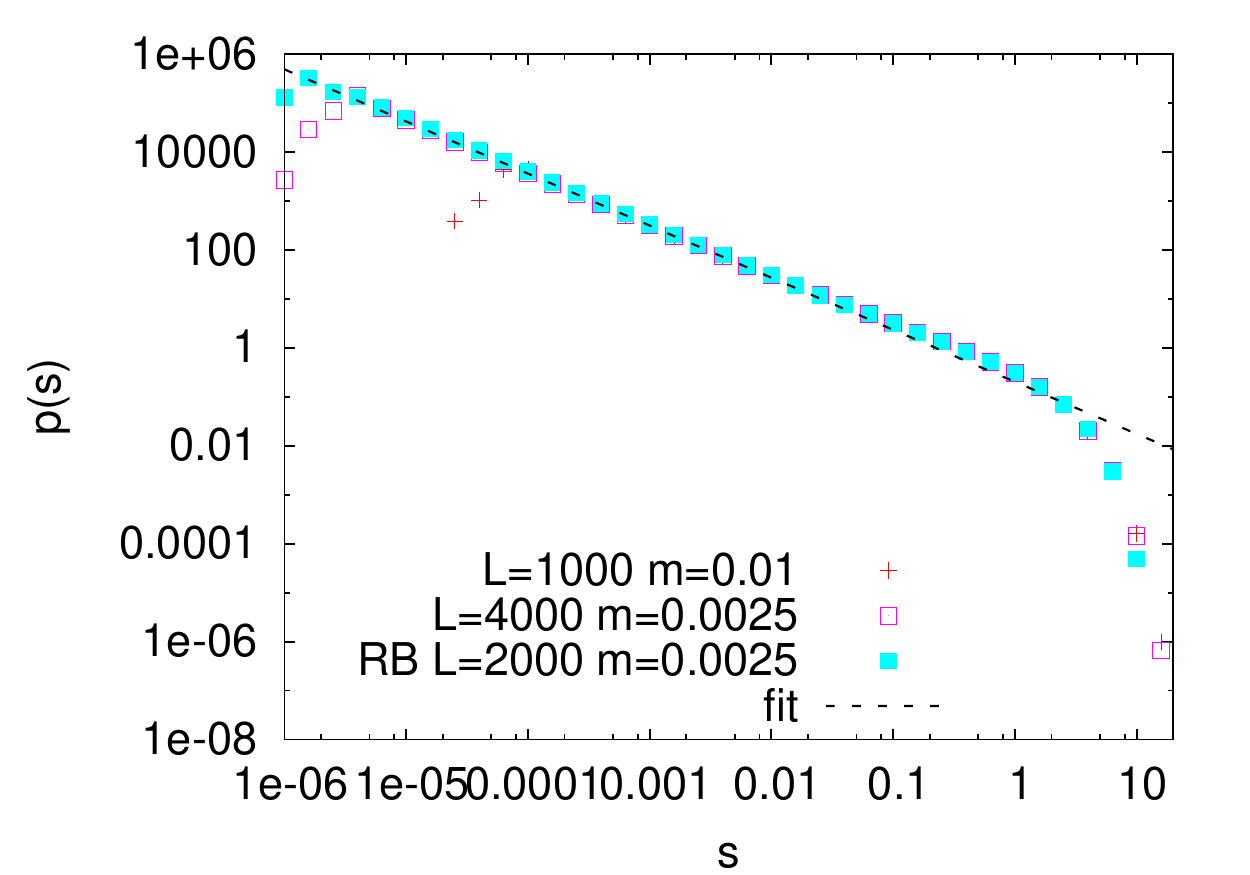}} 
\caption{Random Field and Random Bond (RB) ($d=1$). A fit with a power law gives the exponent  $\tau=1.08 \pm 0.02$. The agreement with Eq.\ (\ref{conjecture}) is discussed in the text. 
}
\label{f:RFconjecture}
\end{figure}

The rescaled avalanche-size distribution can be written as 
\begin{equation}\label{a15}
p (s) = s^{-\tau} f (s)
\end{equation}
where $\tau$ is the avalanche-size exponent, and $f (s)$ the universal
cutoff function \footnote{by universal we mean w.r.t. short scale details. Of course this function
is characteristic of a large scale cutoff provided by a parabolic well}, which tends to a constant for $s \to 0$. For the present model, the only analytical prediction prior to
our work \cite{LeDoussalMiddletonWiese2008,LeDoussalWiese2008c}
concerns the exponent $\tau$, via the above mentioned NF \cite{NarayanDSFisher1993a,NarayanDSFisher1993b} conjecture
\begin{equation} 
\tau = \tau_{\mathrm{conj}} = 2 - \frac{2}{d+\zeta} \ ,
\label{conjecture}
\end{equation}
where $\zeta$ is the roughness exponent at the depinning transition. Exact solution \cite{DSFisher1998} of a mean field toy model of avalanches, which turns out to be related to the famous Galton process \cite{WatsonGalton1875} in genealogy,
gives an exponent $\tau_{\mathrm{MF}}=3/2$. This exponent is also the one
expected if we replace $d=d_{\mathrm{uc}}=4$ in the NF conjecture.  This does however not 
constitute a first-principle calculation starting from the model of the pinned interface.
The latter was only possible using the FRG \cite{LeDoussalMiddletonWiese2008,LeDoussalWiese2008c}. 
The summation of all tree diagrams within the FRG is shown to be asymptotically exact for $d>4$ and
leads to the mean-field prediction \cite{LeDoussalMiddletonWiese2008,LeDoussalWiese2008c} for $\tau$
and for the full rescaled avalanche-size distribution (see below).

We now discuss our numerical results starting with the avalanche-size exponent $\tau$. Note that the data in Fig.\ \ref{f:RFconjecture} contain both random-field and random-bond disorder
and that, as expected from the universality of the depinning fixed point, the results are indistinguishable. Hence in the following we focus on
RF disorder. For $d=1$, a direct power-law fit of our numerical data (see  Fig. \ref{f:RFconjecture}) gives
\begin{equation}\label{taunum}
\tau^{d=1}_{\mathrm{num}}=1.08 \pm 0.02\ .
\end{equation}\begin{figure}[b]
\includegraphics[width=\columnwidth]{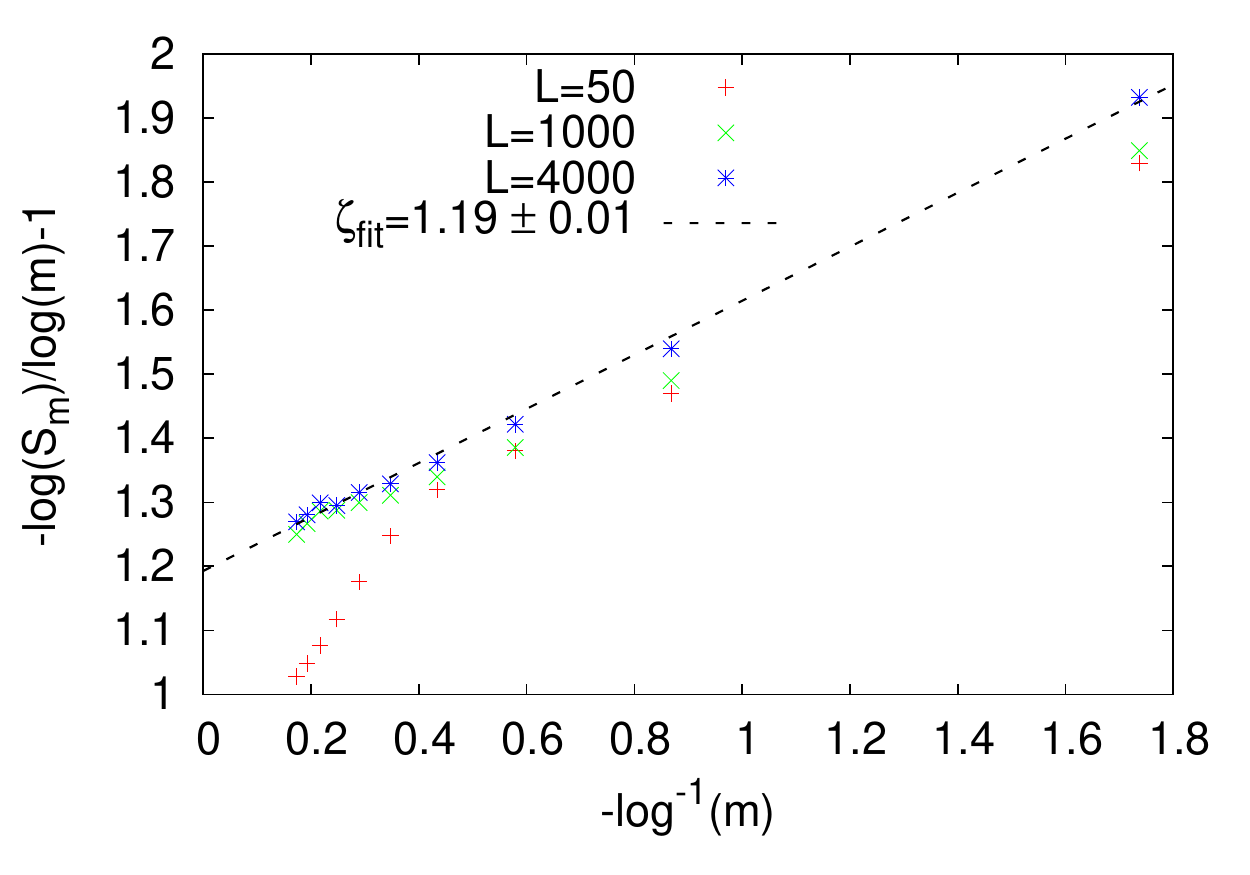}
\caption{Numerical extrapolation of the exponent $\zeta$ to mass $m=0$. We find $\zeta=1.19 \pm 0.01$}\label{figzeta}
\end{figure}This value has to be compared with the conjecture of Eq.~(\ref{conjecture}). The roughness exponent  is known numerically with a good accuracy from system sizes ($L \sim 10^3$) and $m=0$, as $\zeta = 1.26 \pm 0.01$ \cite{RossoHartmannKrauth2002}. This value for $\zeta$ gives $\tau_{\mathrm{conj}}=1.115 \pm 0.005$. Hence the
estimate (\ref{taunum}) is slightly smaller than the value of $\tau$ obtained from the conjecture. There are several possible explanations for this. 

First one notes that although (\ref{taunum}) is extracted from pieces of  $p(s)$ which have already well converged in terms of $m$ and $L$, the resulting window of sizes is limited. Although we took this into account in estimating (\ref{taunum}), we cannot exclude a further small upward shift in the central value as the window size increases. 

Second, we have also measured the {\it effective} $\zeta$ exponent for the sizes and masses used here. 
From measurements of $S_m$ we extract $\zeta=1.19 \pm 0.01$ as can be seen on
Fig.\ \ref{figzeta}. We have checked that comparable estimates can be extracted from the structure factor $S(q)$, as measured also in 
\cite{RossoLeDoussalWiese2006a}, using fits taking into account the mass. Inserting this value for an effective $\zeta$ into
Eq.\ (\ref{conjecture}), this results in an effective value for $\tau_{\mathrm{conj}}=1.086\pm 0.004$, which is in much better agreement with our measured value (\ref{taunum}).

Finally, deviations from the conjecture for the {\it asymptotic} value of $\tau$ are still, strictly speaking, possible, but if they exist they must be around or below the error of $0.02$ in Eq.\ (\ref{taunum}). This does not rule them out since, as discussed in \cite{LeDoussalWiese2008c}, if present they are expected to be small  \footnote{Note that the exact solution in $d=0$ \cite{LeDoussalWiese2008a} shows that the conjecture is valid for RF disorder with sufficiently LR correlations (e.g.\ $\tau=3/2$ and $\zeta=4$ for a Brownian force landscape), but {\it fails} for RF disorder with SR correlations, e.g.\ $\tau=0$ and $\zeta=2$, i.e.\ for the Gumbel class with $\tau < \tau_{\mathrm{conj}}=1$. }. 

Within the FRG  \cite{LeDoussalMiddletonWiese2008,LeDoussalWiese2008c} it is possible to compute the universal scaling function. For $d > 4$ summation of all tree diagrams gives
\begin{equation}
f_{\mathrm{MF}} (s) =\frac{1}{2 \sqrt{\pi}}  e^{-s/4}\ .
\label{meanfield}
\end{equation}
The one-loop FRG calculation gives
\begin{equation}\label{final}
f(s) = \frac{A}{2 \sqrt{\pi}}   \exp\!\left(C \sqrt{s} -
\frac{B}{4} s^\delta\right)\ ,
\end{equation}
 with  exponents
\begin{eqnarray}\label{a64}
&& \tau = \frac{3}{2} + \frac{3}{8} \alpha =  \frac{3}{2} - \frac{1}{8} (1 - \zeta_1) \epsilon 
\\
&& \delta= 1 - \frac{\alpha}{4} = 1 + \frac{1}{12} (1 - \zeta_1) \epsilon\,
\end{eqnarray}
where $\alpha= - \frac{1}{3} (1 - \zeta_1) \epsilon$ and $\zeta_1=1/3$ for the RF class, relevant to the present study. The constants $A$, $B$ and $C$  depend   on  $\epsilon$, and must satisfy  the normalization conditions (\ref{7}), (\ref{8}). At first order in $\epsilon$ they are   $C=- \frac{1}{2} \sqrt{\pi} \alpha$, $B = 1-\alpha(1+\frac{\gamma_{\mathrm{E}}}{4})$, $A = 1+
\frac{1}{8} (2-3 \gamma_{\mathrm{E}} ) \alpha$, $\gamma_{\mathrm{E}}=0.577216$. As usual, the one-loop
results for the exponents $\tau$, $\delta$ and for the  parameters $A$, $B$, and $C$ are exact up to $O(\epsilon^2)$. 

\begin{figure}
\includegraphics[width=\columnwidth]{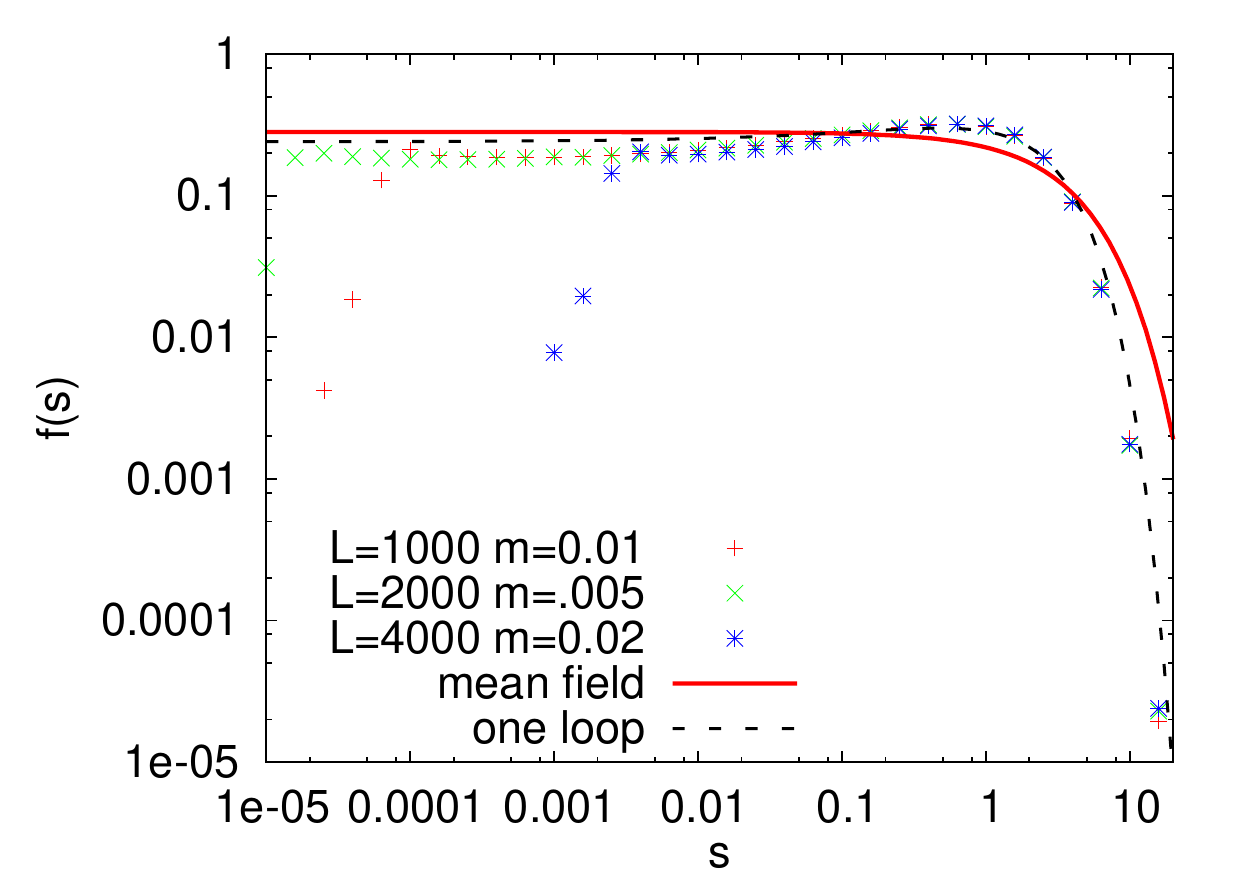} 
\caption{ Random Field ($d=1$). Blow up of the power-law
region. The red solid curve is given by Eq.(\ref{meanfield}) ,  the black dashed line  by
Eq.(\ref{final}), with $A=0.852$, $B=1.56$ and $C=0.56$.}
\label{RFpower}
\includegraphics[width=\columnwidth]{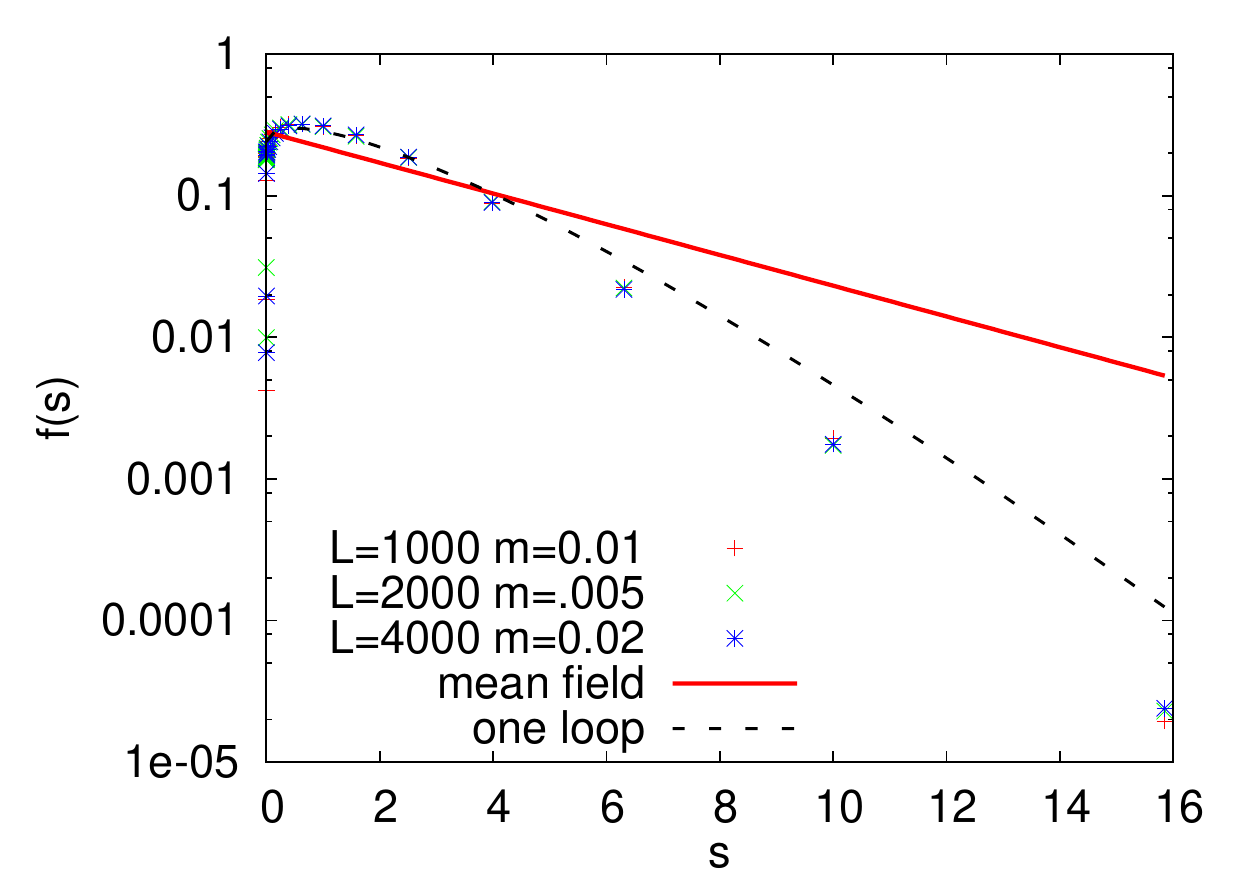}
\caption{Random Field ($d=1$). Blow up of the tail
region. The red solid curve is given by Eq.(\ref{meanfield}) ,  the black dashed line  by
Eq.(\ref{final}), with $A=0.852$, $B=1.56$ and $C=0.56$.}
\label{RFtail}
\end{figure}

To analyze our numerical data for the avalanche-size distribution, we have first computed $\left<S^2\right>$ and $\left<S\right>$ from the data, which allowed to determine numerically the universal (and parameter-free) function $p(s)$ using (\ref{4}) and (\ref{a2}). Hence by construction the numerical data
satisfy conditions (\ref{7}) and (\ref{8}). They are
plotted in Figs.~\ref{RFpower} and \ref{RFtail}, with emphasis either on the power-law region or
on the tail. Note that for the different values of $m$ and $L$ used here, the data have converged, with the exception of the last point for very large avalanches over-suppressed by the finite size of the interface in the smallest samples, and of the region  of very small avalanches, which are cut off  at $s \approx 1/S_m$. 

To compare the numerical data with the mean-field and one-loop predictions, we use two procedures:

In the first procedure we compare directly the cut-off functions $f(s)$, see Figs.~\ref{RFpower} and \ref{RFtail}. They are defined as $f(s):=s^\tau p(s)$ where $\tau$ is respectively $\tau_{\mathrm{num}}=1.08$ for the numerical data, $\tau_{\mathrm{MF}}=3/2$ for the mean-field prediction, and $\tau_{\mathrm{Pade}}=5/4$ for the simplest Pad\'e approximant of the one-loop result, i.e.\ setting $\epsilon=3$ in (\ref{a64}). For $d=1$, due to the large value of $\epsilon=3$, the function $p_ {1\mbox{\scriptsize -loop}}(s)$ with $\tau=5/4$, $\delta=7/6$, $A=5/6-\gamma_{\mathrm{E}}/4$, $B=5/3+\gamma_{\mathrm{E}}/6$ and $C=\sqrt{\pi}/3$ does not have the correct normalization. We chose to introduce two rescaling factors
\begin{equation} \label{proc} 
p(s)= c_1 \,p_{1\mbox{\scriptsize -loop}}(c _2 s)
\end{equation}
in order to enforce the conditions (\ref{7}) and (\ref{8}). This procedure only changes the values of $A$, $B$ and $C$ in a consistent manner, see Figs.~\ref{RFpower} and \ref{RFtail}. Note that even though only  $f(s)$ is plotted, the chosen value of $\tau$ changes the values of $A,B,C$ via the normalization conditions, hence must be discussed accordingly.

\begin{figure}[t]
{\includegraphics[width=\columnwidth]{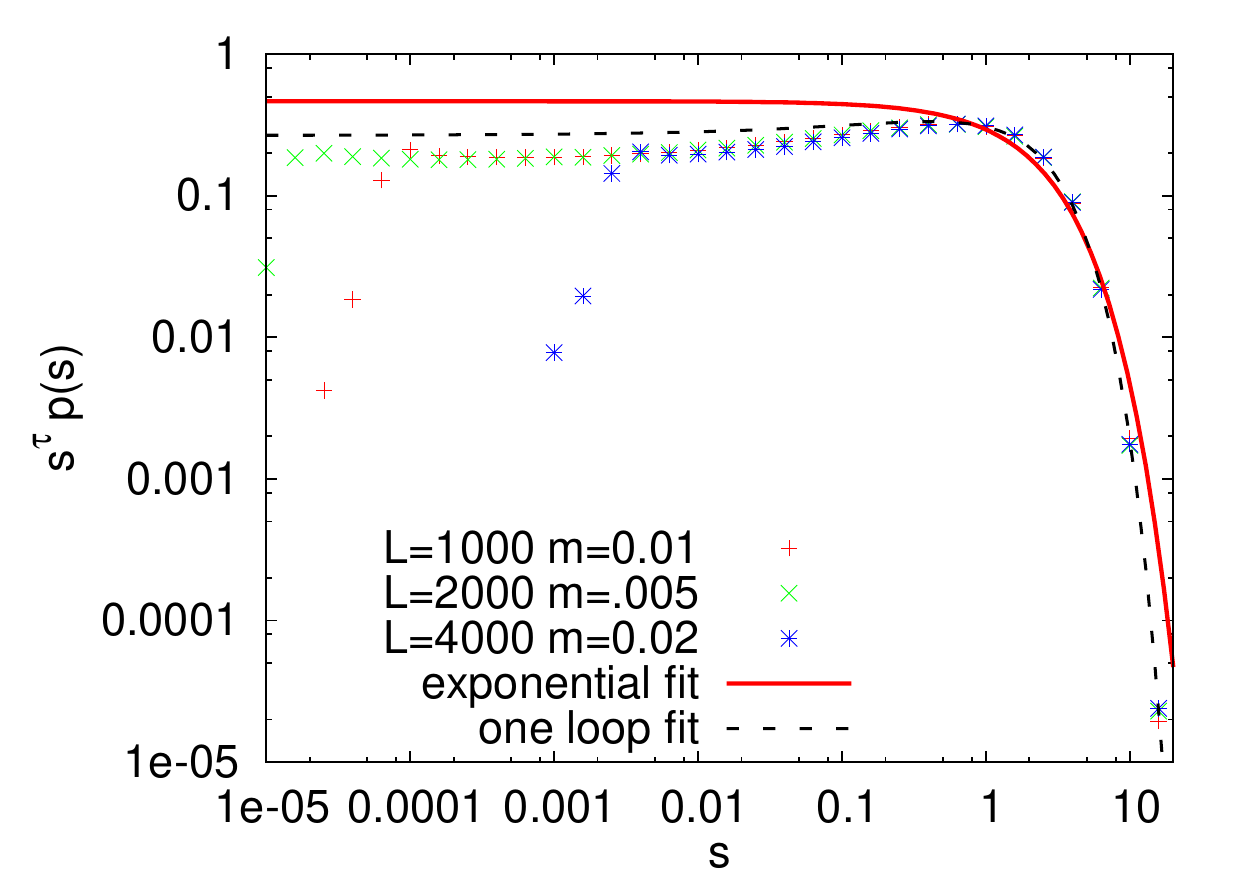}} 
\caption{ Random Field ($d=1$). Blow up of the power-law
region. The red solid curve is given by Eq.(\ref{kayexp}) ,  the black dashed line  by
Eq.(\ref{final}), with $A=0.947$, $B=1.871$ and $C=0.606$.}
\label{RFpowerkay}
{\includegraphics[width=\columnwidth]{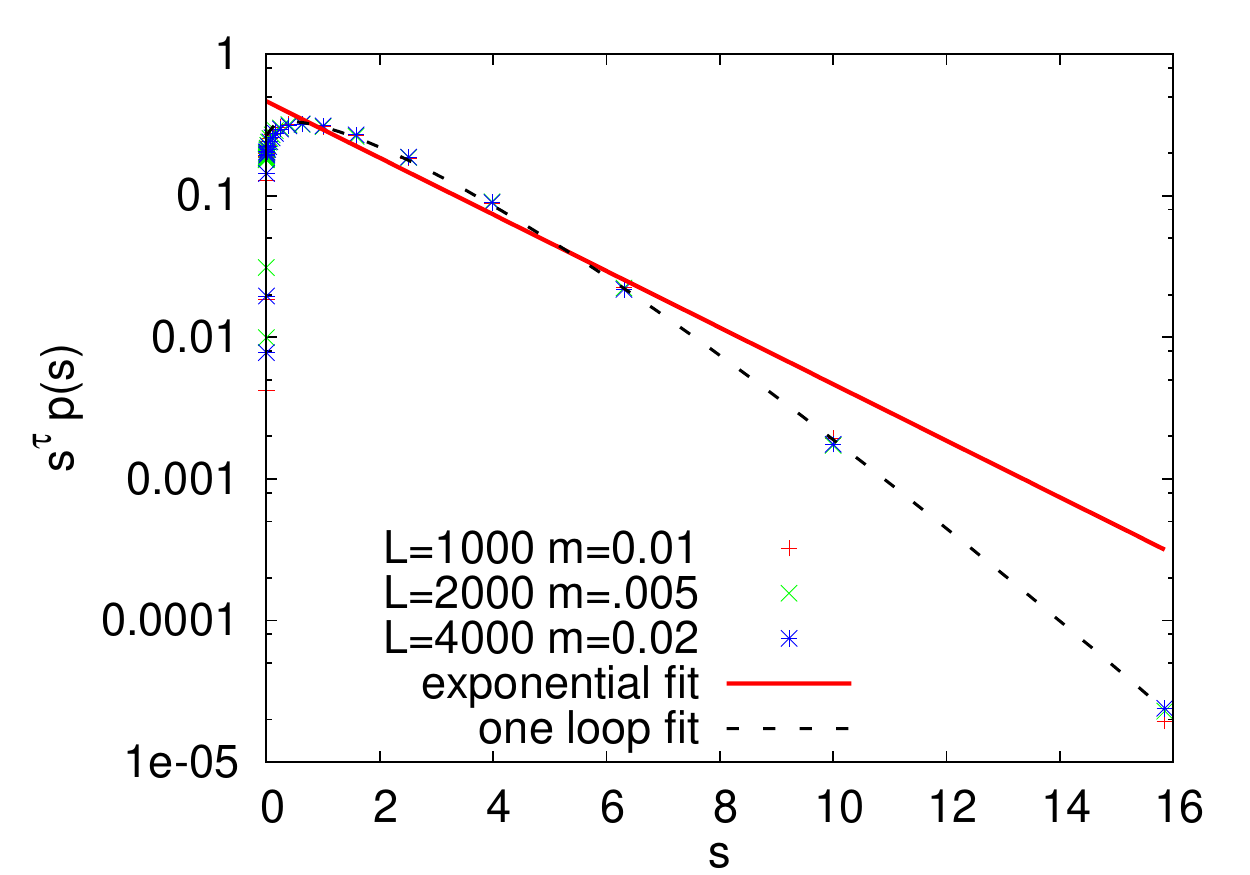}}
\caption{Random Field ($d=1$). Blow up of the tail
region. The red solid curve is given by Eq.(\ref{kayexp}) ,  the black dashed line  by
Eq.(\ref{final}), with $A=0.947$, $B=1.871$ and $C=0.606$.}
\label{RFtailkay}
\end{figure}

A second approach, shown in Figs.~\ref{RFpowerkay} and \ref{RFtailkay}, consists in fitting {\it the same numerical curves} as in Figs.~\ref{RFpower} and \ref{RFtail}, with either (i) an exponential function (``exponential fit'') or (ii) the one-loop function (``fit one loop''), but using the numerically obtained exponent $\tau=\tau_{\mathrm{num}}=1.08$. The exponential fit reads
\begin{equation}\label{kayexp}
p (s) s^{\tau }  = \frac{\left(1-\frac{\tau }{2}\right)^{2-\tau
   }}{\Gamma (2-\tau )} \exp\! \left(\left(-1+\frac{\tau }{2}\right) s\right)\ .
\end{equation}
All coefficients are determined as a function of $\tau$ by the normalization conditions (\ref{7}) and (\ref{8}).
Note that this exponential fit is mostly a guide to emphasize the sub-exponential tail apparent in the data. 
Similarly, for the one-loop
fit we adopt the procedure described in the previous paragraph, with  $\tau=\tau_{\mathrm{num}}$ everywhere instead of the one-loop Pad\'e value $\tau=5/4$. We expect this fit to be less sensitive to the lack of precision in the one-loop estimate of $\tau$ for the large value of $\epsilon=3$ relevant here, and to better capture the tail region. This is indeed what is found, see Fig.\ \ref{RFtailkay}. It confirms the sub-exponential tail exponent $\delta \approx 7/6$ to a rather good precision. We stress that our procedure is {\it not a fit} using $A,B,C$ as fit parameters, but that {\em all} parameters are specified by the one-loop prediction.

\begin{figure}
{\includegraphics[width=\columnwidth]{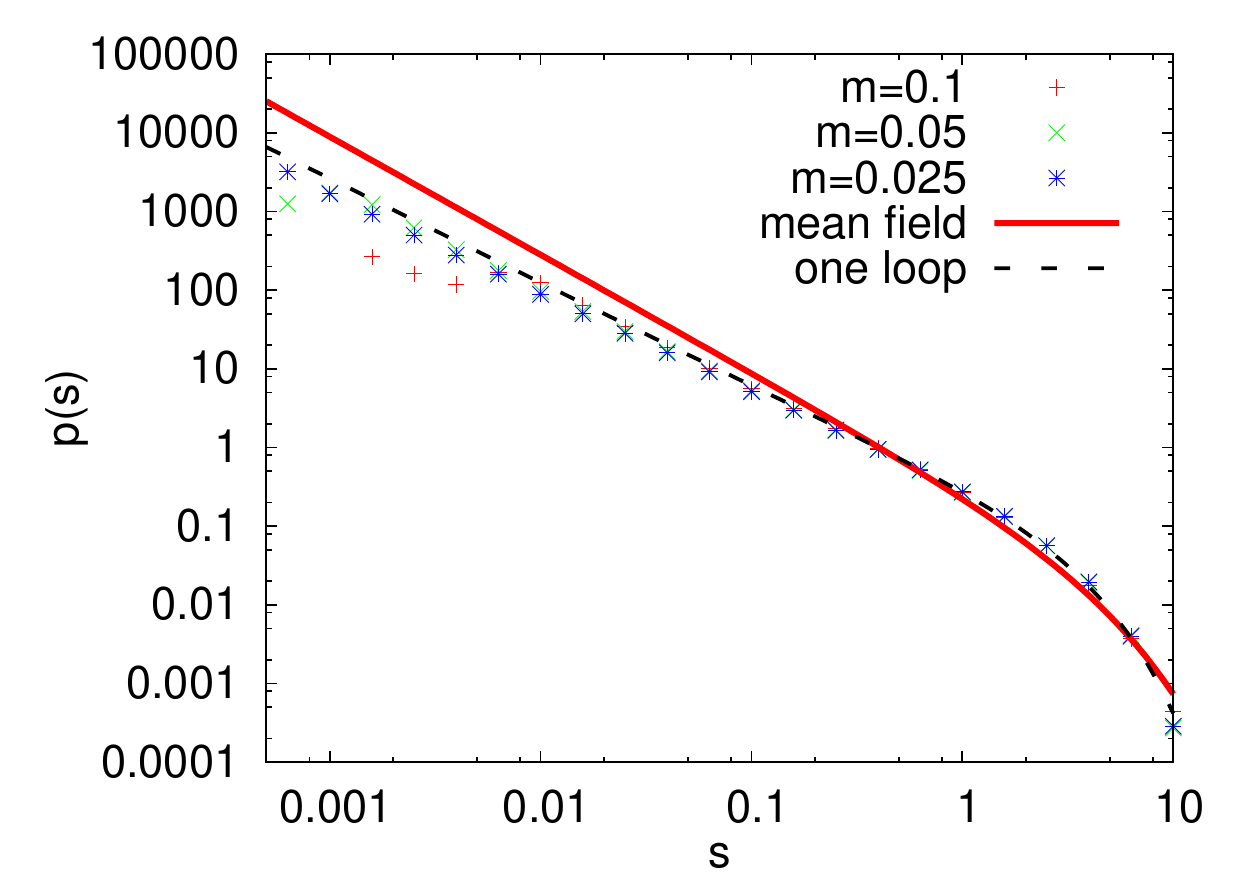}} 
\caption{Random Field ($d=2$, $L=100$ for $m=0.1$, $0.05$; $L=200$ for $m=0.025$). A fit with a power law gives access to the exponent  $\tau=1.3 \pm 0.01$. The comparison with Eq.(\ref{conjecture}) is discussed in the text.
}
\label{f:RFconjecture2d}
\end{figure}
\medskip

We now turn to a 2-dimensional interface.
The universal function $p(s)$ for $d=2$ and RF disorder is plotted on Fig.\ \ref{f:RFconjecture2d}. 
From a direct power-law fit, we find
\begin{equation}\label{20}
\tau_{\mathrm{num}}^{d=2}=1.3 \pm 0.02\ .
\end{equation}
This value has to be compared with the conjecture of Eq.~(\ref{conjecture}). The roughness exponent at the depinning transition is known numerically as $\zeta_{\mathrm{num}}^{d=2} = 0.753 \pm 0.002$ \cite{RossoHartmannKrauth2002}, which gives $\tau_{\mathrm{conj}}^{d=2}=1.2735 \pm 0.0005$.
Although our value (\ref{20}) of $\tau$ is compatible with the conjecture, the precision is insufficient to conclude on possible small deviations from the latter. The mean-field and one-loop predictions discussed above are plotted for comparison, using the simplest one-loop Pad\'e approximant, i.e.\ $\alpha=-4/9$, $\tau=4/3$, $\delta=10/9$.  After the above described procedure (\ref{proc}) using the normalization conditions (\ref{7}) and (\ref{8}) this led to  the values $A=0.92$, $B=1.416$ and $C=0.383$.


%



\section{Universal moment ratios $r_n$}
\label{s:5}
\begin{figure}
{\includegraphics[width=1.\columnwidth]{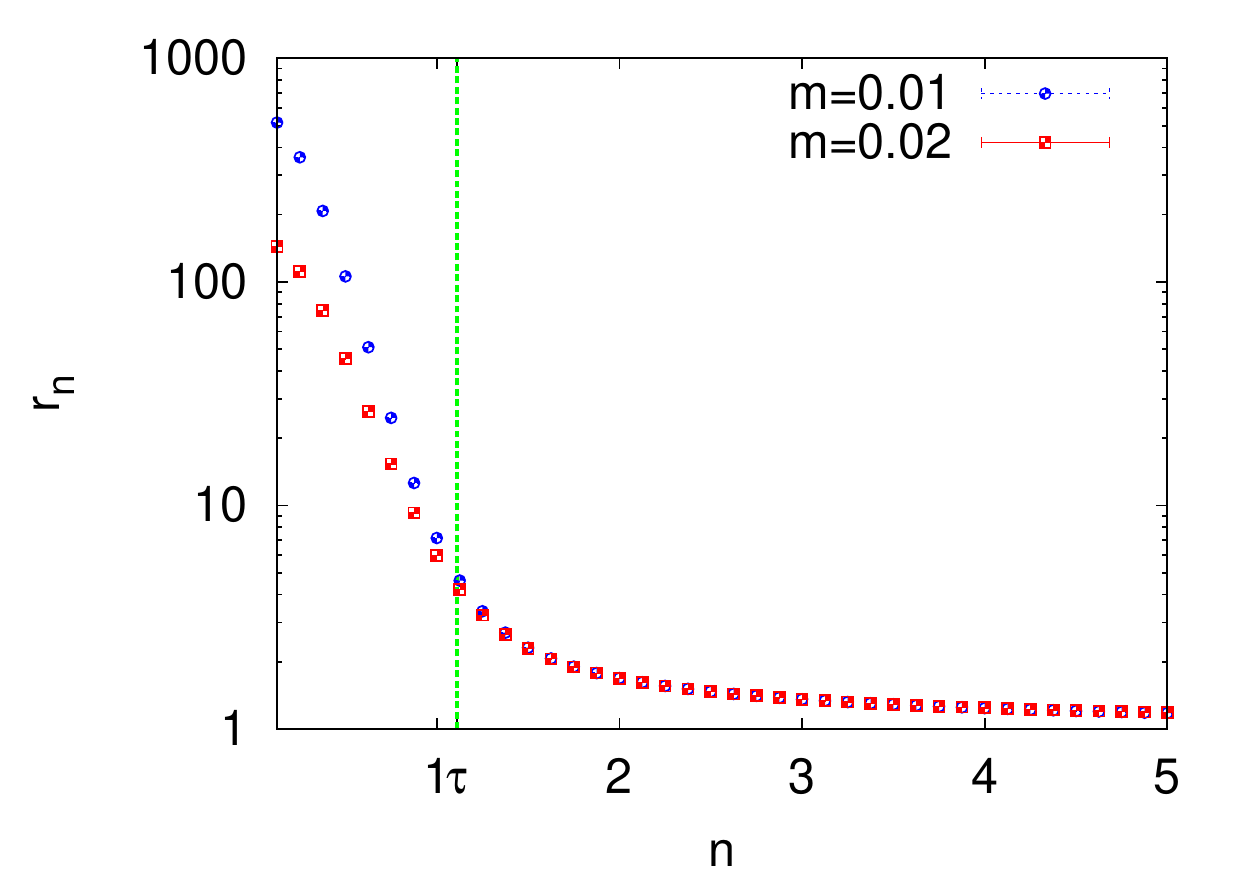}} 
\caption{Numerical values of the ratios $r_n$ for bare random-field disorder ($d=1$, $L=2000$),
and $m=0.02$, $m=0.01$. 
For $m=0$ a pole is expected for $n=\tau$. 
For $n<\tau$ the divergence manifests itself through the data's dependence on $m$ . 
}
\label{f:RFrntotal}
\includegraphics[width=1.\columnwidth]{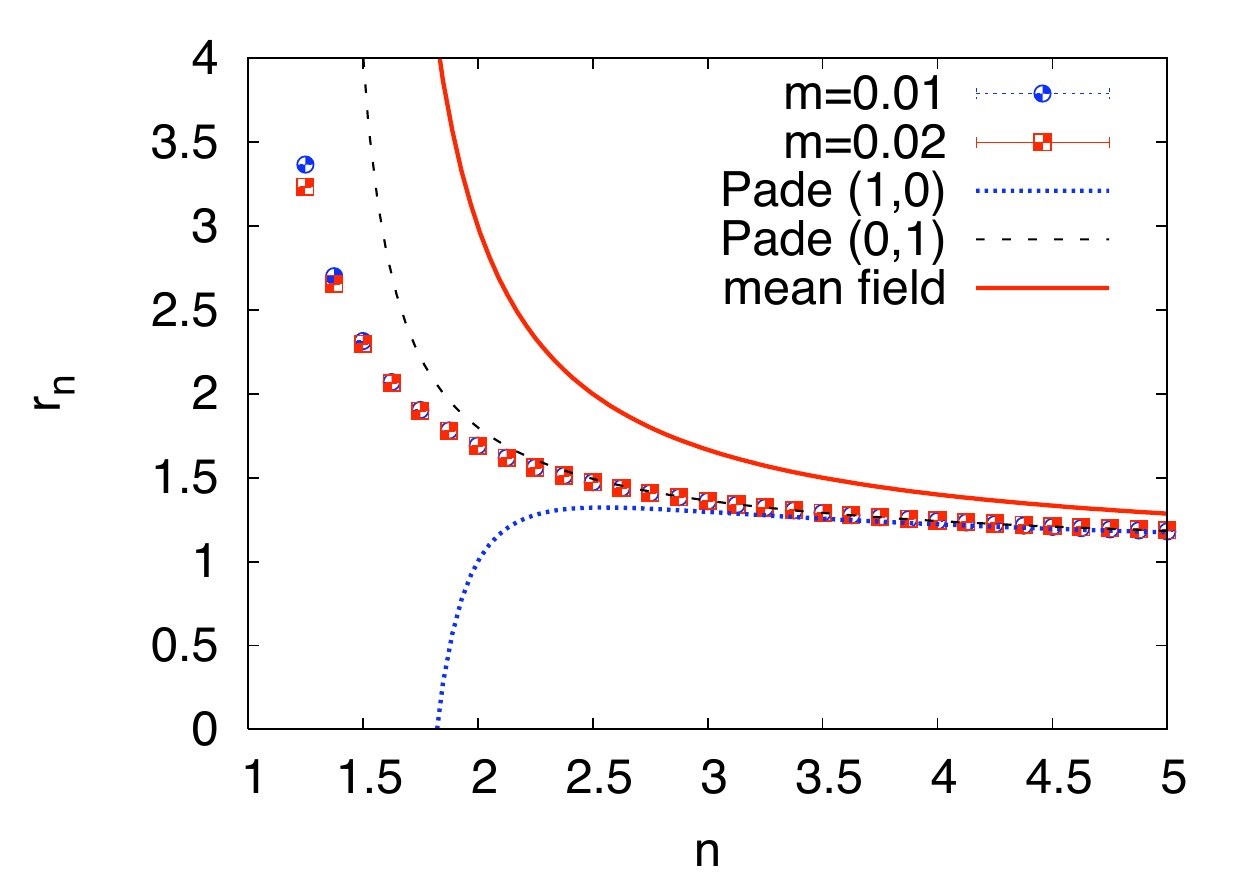}
\caption{Random Field ($d=1$, $L=2000$). Moment ratios $r_n$:
comparison between
numerics and analytic predictions}
\label{f:RFrn}
\end{figure}
\begin{figure}
\includegraphics[width=1.\columnwidth]{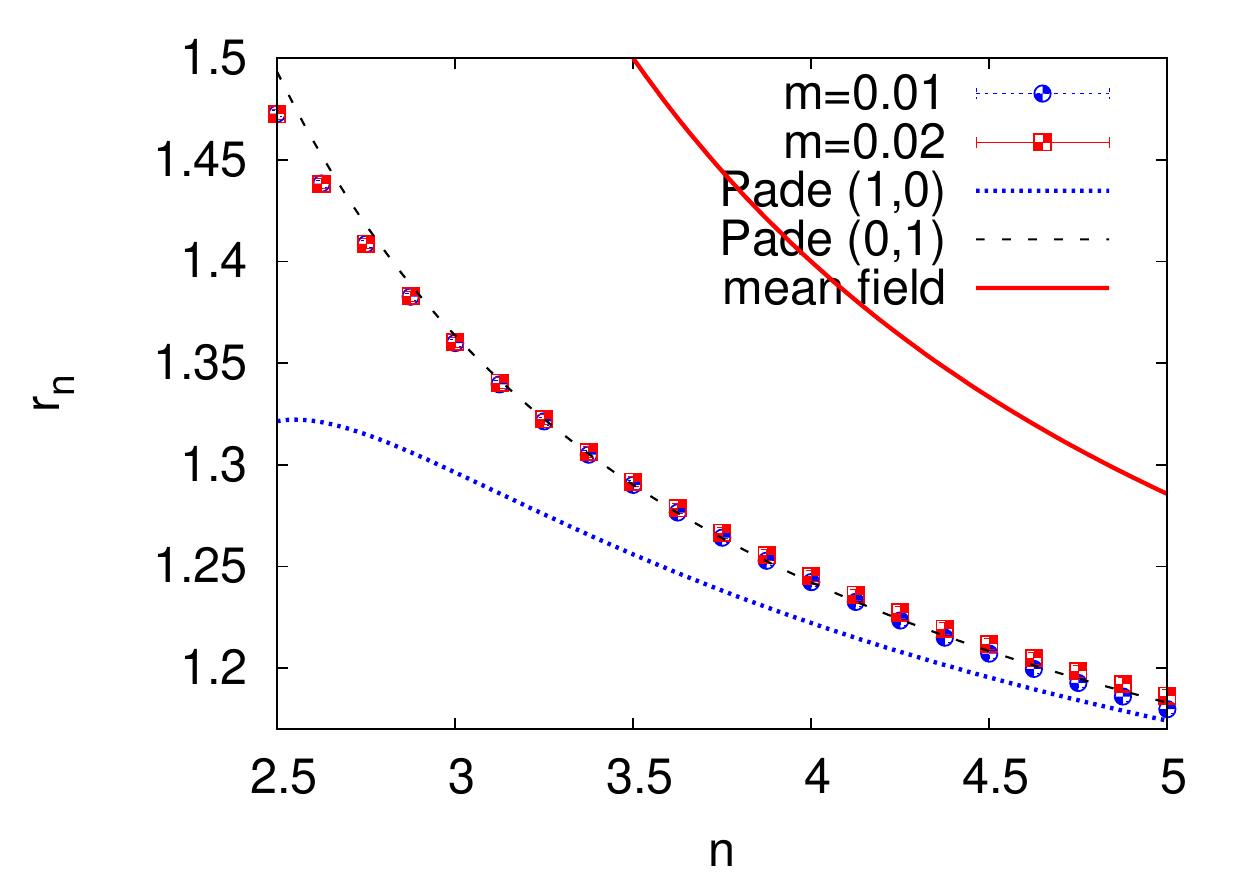} 
\caption{
Random Field ($d=1$, $L=2000$). Moment ratios $r_n$: blow up of the tail behavior. The mean
field behavior is given by Eq.(\ref{MF}). The Pad\'e $(1,0)$ by
Eq(\ref{pade10}) and the Pad\'e $(0,1)$ is also plotted. }
\label{f:RFrntail}
%
{\includegraphics[width=1.\columnwidth]{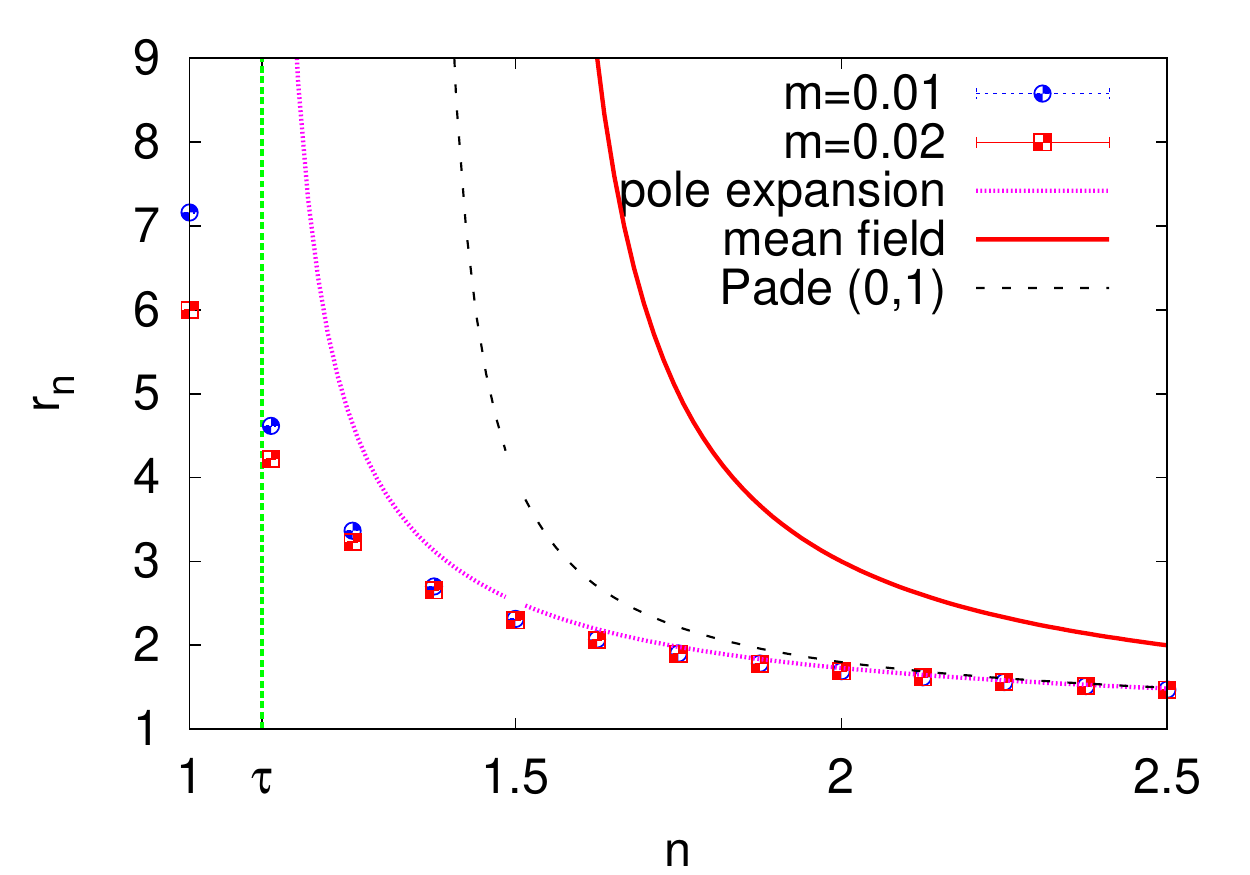}} 
\caption{
Random Field ($d=1$, $L=2000$).  Moment ratios $r_n$: blow
up around the pole and comparison with Eq.~(\ref{poleexp}).}
\label{f:RFrnpole}
\end{figure}
Important universal quantities characterizing the avalanche statistics are the following universal ratios of avalanche-size moments:
\begin{eqnarray}
 r_n:&=& \frac{\langle S^{n+1}\rangle \langle S^{n-1}\rangle}{\langle S_n\rangle^2} \nonumber \\
&=&  \frac{\langle s^{n+1}\rangle_p \langle s^{n-1}\rangle_p}{\langle s_n\rangle_p^2}\ .
\end{eqnarray}
Here $n$ can be non-integer. As shown in \cite{LeDoussalWiese2008c} all non-universal scales disappear
in the ratios $r_n$. Our numerical findings are summarized in
Fig.\ \ref{f:RFrntotal}. The pole expected
at $n=\tau \approx 1.08$ in the limit of infinite $S_{m}/S_{\mathrm{min}}$
manifests itself in a non-convergence of the numerical data upon
lowering $m$. This is an independent method for calculating $\tau$.

We now compare to the FRG calculation \cite{LeDoussalWiese2008c}.
The function $r_n$ can be evaluated in an $\epsilon=4-d$ expansion.
At the mean-field level ($\epsilon=0$)
\begin{equation}
r_n^{0}=\frac{2 n -1}{2 n -3}\ ,
\label{MF} 
\end{equation}
and a pole is found for $n=\tau_{\mathrm{MF}}=3/2$.

The one-loop $\epsilon$ expansion leads to the following expression \cite{LeDoussalWiese2008c}:
\begin{equation}
r_n^{\mathrm{\mbox{\scriptsize 1-loop}}}=r_n^{\mathrm{MF}}  -\frac{\epsilon}{3} (1 -\zeta_1)\frac{n \Gamma(n-\frac{3}{2} )+\sqrt{\pi} \Gamma(n-1)}{(2 n -3)^2\Gamma(n-\frac{3}{2})}\ ,
\label{pade10}
\end{equation}
where $\zeta_1=1/3$ for RF. This expression corresponds to the Pad\'e
$(1,0)$ in the $\epsilon$-expansion; we  also use  the  Pad\'e
$(0,1)$. The comparison with the data is shown on Fig.~\ref{f:RFrn}. 
For the large-moment region, a blow-up is shown on Fig.~\ref{f:RFrntail}. 
The agreement of the data with the two one-loop Pad\'e approximants, as compared
to mean field, is quite striking.

\renewcommand{\topfraction}{0.8}
\begin{figure*}[t]\includegraphics[width=1.02\columnwidth]{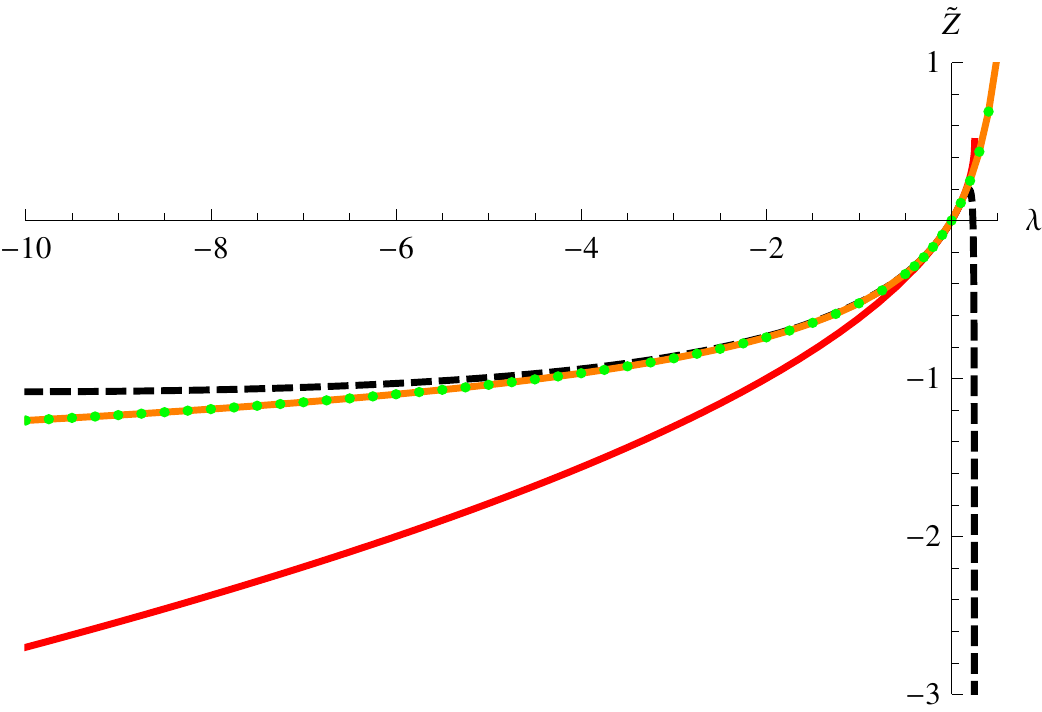}\hfill
\includegraphics[width=1.02\columnwidth]{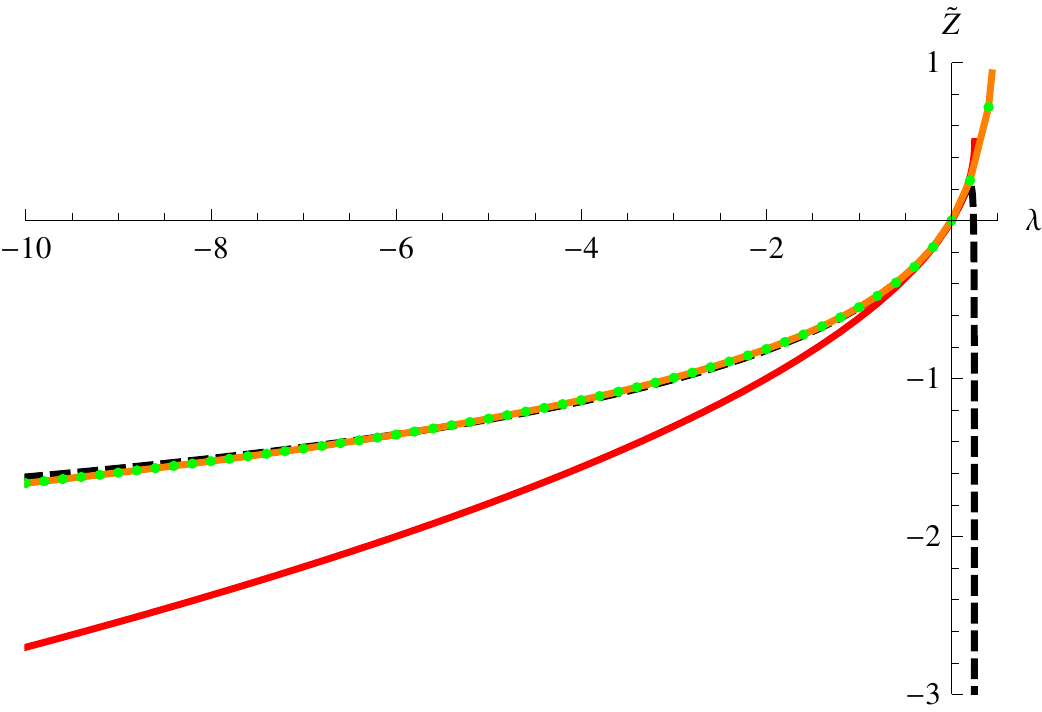}
\caption{The characteristic function $\tilde Z (\lambda)$ for $d=1$ (left) and $d=2$ (right). Mean field
(solid red/grey) from Eq.~(\ref{a9}); one-loop from Eq.~(\ref{a61}) (dashed black); and
numerical results for $L=4000, m=0.0005$ (solid orange/green dots). The singularity in (\ref{a9}) and (\ref{a61}) for $\lambda= 1/2$ (indicated by a vertical dotted line) is smoothed out in the numerics, see figure \ref{a11b} for details.}
\label{a11}
\includegraphics[width=1.02\columnwidth]{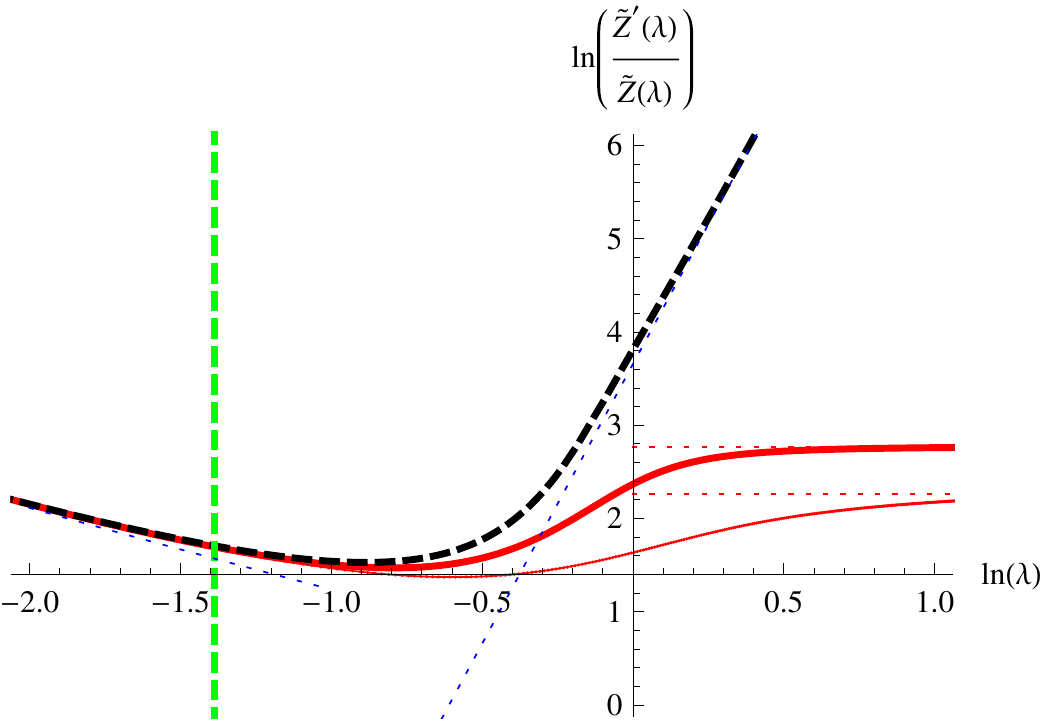}\hfill
\includegraphics[width=1.02\columnwidth]{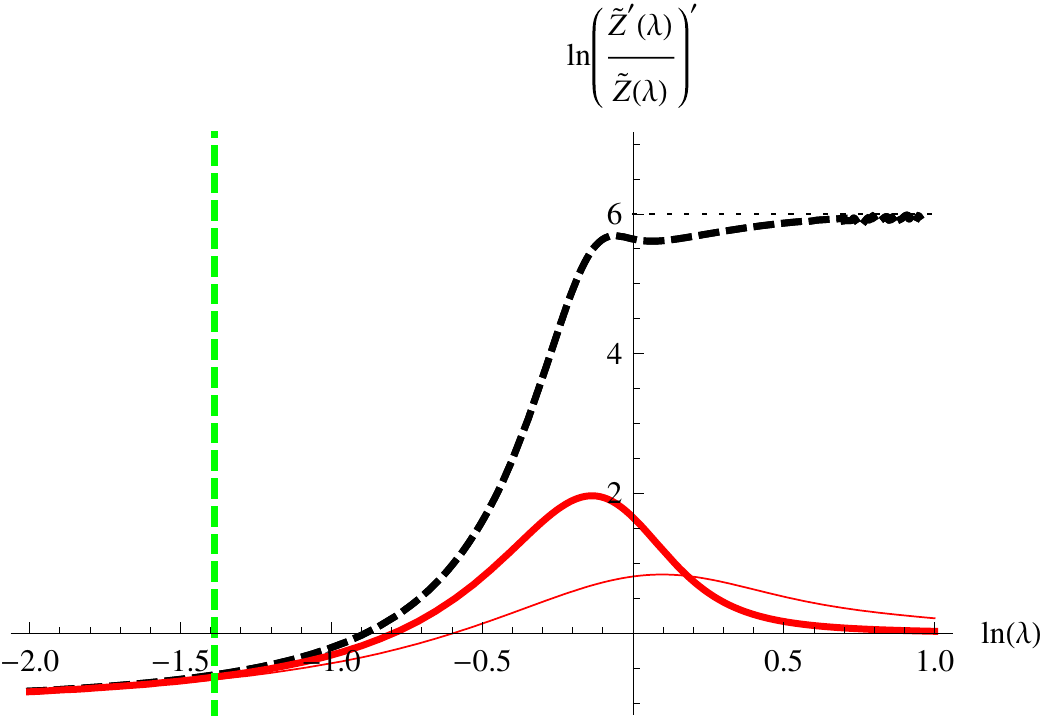}
\caption{Left: Log-log plot of $\partial_\lambda \ln \tilde Z(\lambda)$ versus $\lambda$. For large $\lambda$, we expect from Eq.~(\ref{final}) that the analytic result (black, thick, dashed) obtained by integrating Eq.\ (\ref{a8}) numerically, has slope $1/(\delta-1)$ (blue dotted line is this asymptotics). The numerical results are for $L=4000$ and $m=0.02$ (thick red) and $m=0.00125$ (thin red).
For large $\lambda$, $\tilde Z_{\mathrm{num}}(\lambda) \approx \frac{S_{m}}{\left< S \right>}  \frac{1}{N} \exp(\lambda S_{\mathrm{max}})$, where $S_{\mathrm{max}}$ is the largest avalanche encountered in the simulation, and the curve saturates (dotted red lines). The larger $S_{\mathrm{max}}/S_m$, the better the data. For $m=0.02$ (thick red) this ratio is 16, whereas for $m=0.00125$ (thin red) it is 10. The vertical dashed green line indicates the location $\lambda =1/4$ of the singularity in $\tilde Z(\lambda)$ at the mean-field level. 
Right: Slope of the function left, i.e. the effective exponent $1/(\delta-1)$. One sees that the effective exponent increases with increasing $S_{\mathrm{max}}/S_m$. Our data, which clearly have not converged in terms of $S_{\mathrm{max}}/S_m$, allow to estimate $1< \delta\le \frac3 2$ from the maximum slope.}
\label{a11b}
\end{figure*}
However, both Pad\'es break down close to $n=\tau$. We
give another useful form for comparison to numerics. The idea is to isolate the simple pole  which occurs
in any dimension, as
\begin{equation}
r_n=\frac{A_d}{n-\tau} +B_{n,d}\ .
\label{poleexp}
\end{equation}
Up to $O(\epsilon^2)$ corrections  \cite{LeDoussalWiese2008c}
\begin{eqnarray}\label{a16}
A_d&=&1-\frac{1+\pi}{12}(1-\zeta_1)\epsilon \\ 
B_{n,d}&=&1+
\frac{\pi \Gamma(n-\frac{1}{2} )-\sqrt{\pi} \Gamma(n-1)
}{6 (2 n -3)\Gamma(n-\frac{1}{2})}(1-\zeta_1)\epsilon\ .\qquad 
\end{eqnarray}
In 
Fig. \ref{f:RFrnpole} this formula is plotted setting $\epsilon=3$. It shows that it works quite well,
even close to the pole at $n=\tau$.




%

\section{The characteristic function $\tilde Z(\lambda)$}
\label{s:6}

It is useful to define a generating function of exponential moments, i.e.\ the
characteristic function of the avalanche-size probability. Using the definitions (\ref{4}) and (\ref{a4}), 
we define the normalized generating function $\tilde Z
(\lambda)$
\begin{equation}\label{a6}
\tilde Z (\lambda) := \frac{S_{m}}{\left< S \right>}  \frac{1}{N} \sum_{i=1}^{N} \left[ \rme^{\lambda S_{i}/S_{m}}-1 \right]
\end{equation}
By construction, 
$\tilde Z (\lambda) = \lambda + {\lambda^{2}}+ \dotsb \ .$ Since large negative $\lambda$ probe small avalanches, it is expected to be universal for $\lambda \gg - 1/S_{\mathrm{min}}$. In the universal range, its relation to $p (s)$ is 
\begin{equation}\label{a8}
\tilde Z (\lambda) = \int_{0}^{\infty} \rmd s \, p (s)\left(\rme^{\lambda s}-1 \right)\ .
\end{equation}
It has been calculated in
\cite{LeDoussalMiddletonWiese2008,LeDoussalWiese2008c}
at the mean-field level: 

\pagebreak
\begin{equation}\label{a9}
\tilde Z_{\mathrm{MF}} (\lambda) = \frac{1}{2} \left(1-\sqrt{1-4 \lambda }\right)\ .
\end{equation}At 1-loop order, it reads
\cite{LeDoussalMiddletonWiese2008,LeDoussalWiese2008c} \begin{widetext}
\begin{equation}\label{a61}
\tilde Z_{\mbox{\scriptsize 1-loop}} (\lambda) = \frac{1}{2} \left(1-\sqrt{1-4 \lambda }\right)+\frac{\left(\left(3
\lambda +\sqrt{1-4 \lambda }-1\right) \log (1-4 \lambda )-2 \left(2
\lambda +\sqrt{1-4 \lambda }-1\right)\right) \alpha }{4 \sqrt{1-4
\lambda }}+O\left(\alpha ^2\right)\ .
\end{equation}
\end{widetext}
where, as above, 
$
\alpha = -\frac{1-\zeta_{1}}{3}\epsilon  = -\frac{2}{9} \epsilon
$.
The comparison between theory and
numerical data is presented on Fig.\ \ref{a11}, both for  $d=1$, and  $d=2$. 
In these figures we have plotted (\ref{a61}), discarding the term $O(\alpha^2)$ and
setting $\epsilon =3$ and $\epsilon=2$ respectively. 
The plots show that the simplest extrapolation of the 1-loop correction is extremely good in
calculating the behavior even for large negative $\lambda$, as was
already observed in the static case in \cite{LeDoussalMiddletonWiese2008}.  
It would be interesting to compare $\tilde Z(\lambda)$ for both cases numerically.

For large $\lambda$, $\tilde Z(\lambda)$ is dominated by the largest avalanche $S_{\mathrm{max}}$. If $S_{\mathrm{max}}/S_m\gg 1$, then the tail-exponent $\delta$ can in principle be extracted from the derivative of $\ln \tilde Z(\lambda)$, i.e. $\partial_\lambda \ln \tilde Z(\lambda) \sim \lambda^{1/(\delta-1)}$ in some window of $\lambda$ before it eventually saturates to a constant $\sim S_{\mathrm{max}}$ at larger $\lambda$. Our data, which are plotted on Fig.~\ref{a11}, are not yet converged in terms of the ratio $S_{\mathrm{max}}/S_m$, but are sufficient to give the bound $1<\delta\le \frac32$.
One finally notes that although mean field works better for $d=2$ than for $d=1$, the 1-loop corrections are necessary to account for the numerical data.

\section{Local avalanche-size distribution}

\label{sec:local}
\begin{figure}[b]
{\includegraphics[width=\columnwidth]{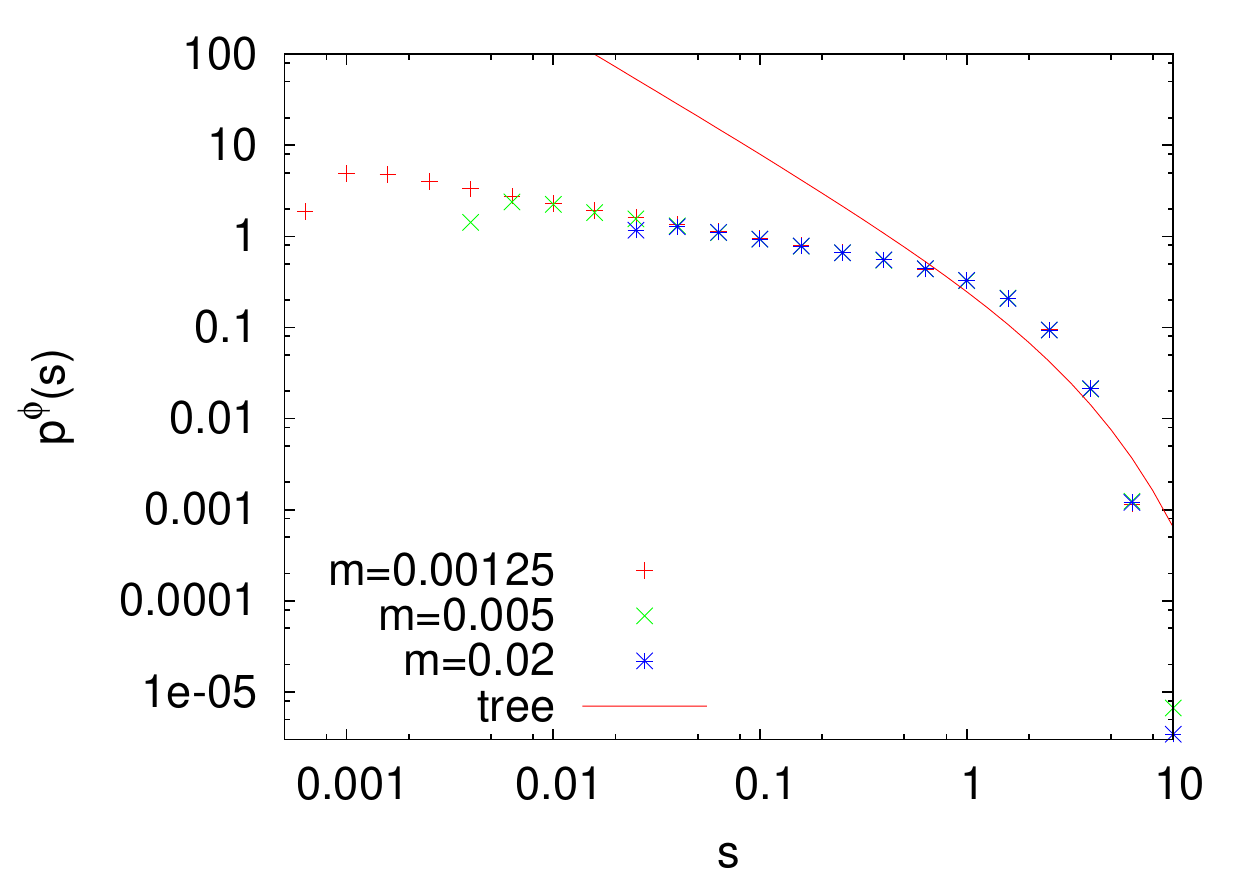}} 
\caption{Random Field ($d=1$, $d_\phi=0$, $L=4000$). From the fit we get $\tau_\phi=0.39 \pm 0.01$.}
\label{f:RFconjecturephi}
\end{figure}
\begin{figure}[bp]
\includegraphics[width=\columnwidth]{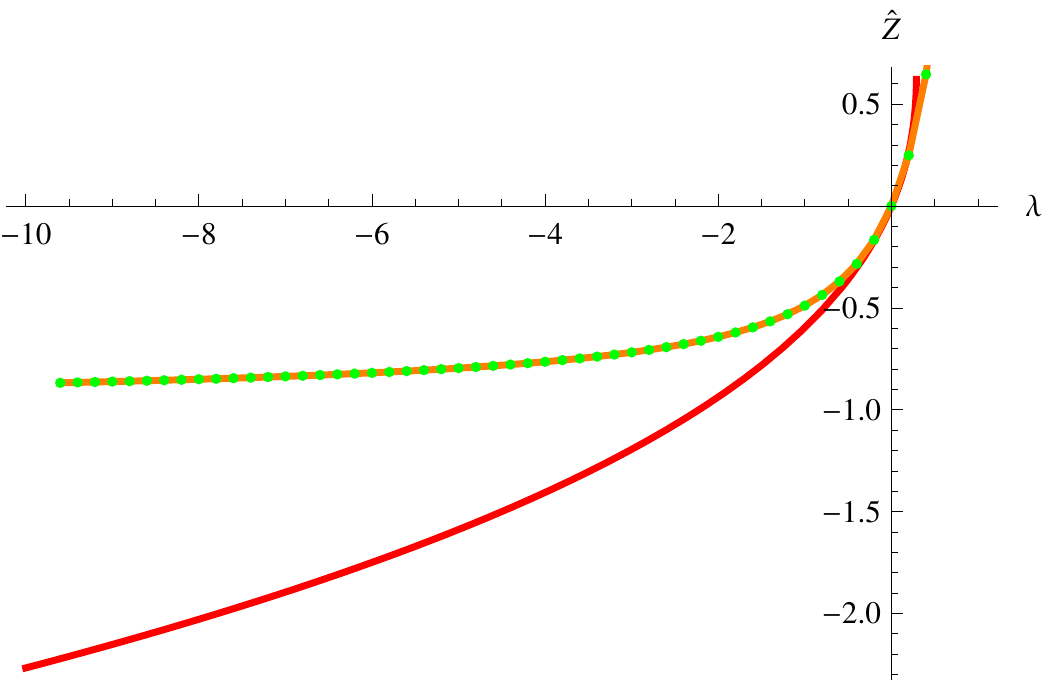}
\caption{$\hat Z(\lambda)$ both at the tree-level (solid/red), and numerically (green/orange dots), for RF disorder, $m = 0.00125$, $d=1$, $d_\phi=0$.}
\label{f:Zhat}
\end{figure}
\begin{figure}[htbp]
\includegraphics[width=\columnwidth]{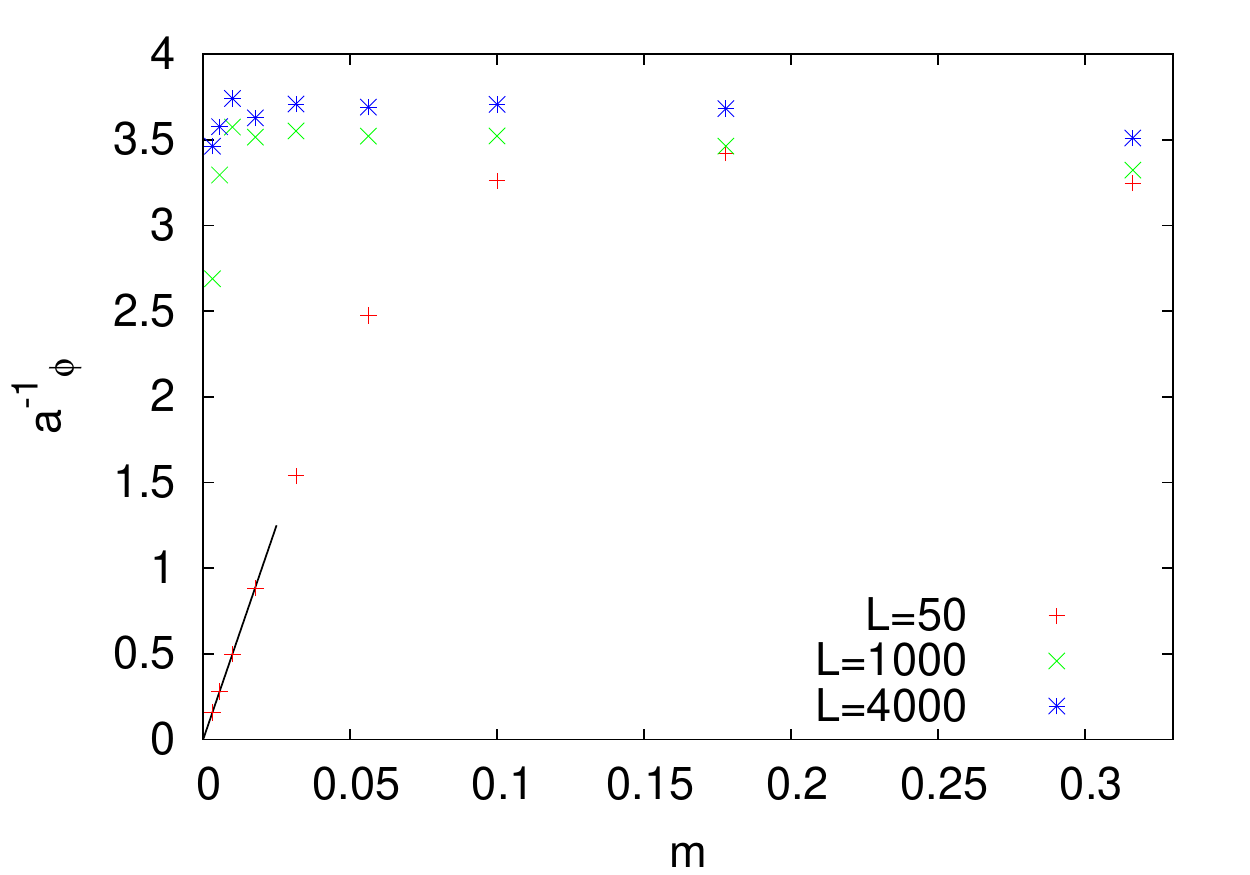}
\caption{$1/a_\phi$ as defined in equation (\ref{aphi}) and discussed below. For $m\ll 5/L$, $1/a_\phi \approx Lm$.}
\label{f:a}
\end{figure}



In \cite{LeDoussalWiese2008c}, we have considered the following definition of the size of 
a local avalanche $S_\phi$:
\begin{equation} \label{resphi}
S_\phi = \int \rmd^d x\, \phi(x) [u_{w_1}(x) - u_{w_0}(x)]\ . 
\end{equation}
Here we also define:
\begin{equation}\label{}
S_m^{\phi} := \frac{\left<S_\phi^2\right>}{2 \left<S_\phi\right>}
\end{equation}
Of particular interest is the cross-section with a co-dimension one hyper-plane i.e.\ $\phi(x)=m^{-1} \delta(x_1)$,
or more generally, with a co-dimension $d'$ subspace. This cross-section has dimension $d_\phi=d-d'$. 
We have chosen the factor of $m$ in the definition of $\phi$ such that $S_m$ and $S_m^\phi$ both scale as $m^{-d- \zeta}$. For $d=1$ we consider a
point, i.e.\ $d'=1$, $d_\phi=0$. Note that we always chose the factor of $m$ in the definition of $\phi$ (see above) such that $S_m$ and $S_m^\phi$ both scale as $m^{-d- \zeta}$. 

For a more convenient comparison with numerics, we adopt a slightly different normalization as in Ref.~\cite{LeDoussalWiese2008c}, and chose to normalize using $S_m^\phi$ rather than $S_m$. We estimate below the ratio $a_\phi=S_m^\phi/S_m$ which allows to go from one set of definitions to the other. Hence the (normalized) local avalanche-size distribution $P^\phi(S^\phi)$ is expected to
take the form
\begin{equation}\label{a2new}
P^\phi (S^\phi) = \frac{\left< S^\phi \right>}{(S^\phi_m)^2}\, p_\phi\! \left(\frac{S^\phi}{S^\phi_{m}}\right)\ .  
\end{equation}
where the universal function $p_\phi(x)=a_\phi^2 \frac{S_m}{\left< S_\phi \right>} p^\phi(a_\phi x)$ in terms of the one defined in \cite{LeDoussalWiese2008c} and called $p^\phi$ there. By construction $p_\phi(x)$ satisfies the normalizations (\ref{7}) and (\ref{8}).
Similarly we define the generating function as 
\begin{equation}
\hat Z(\lambda): = \frac{S_{m}^\phi}{\left< S^\phi \right>}\frac{1}{N} \sum_{i=1}^{N} \left[\rme^{\lambda S^\phi_{i}/S_{m}^\phi}-1\right]\ ,
\end{equation}
which reads $\hat Z(\lambda)=a_\phi \tilde Z^\phi(\lambda/a_\phi)$ in terms of the one defined in \cite{LeDoussalWiese2008c}. 

At present time we have only three analytical results available to compare the numerical data on local avalanche-size distributions. First the conjecture put forward in \cite{LeDoussalWiese2008c} and which generalizes (\ref{conjecture}) reads:
\begin{equation}\label{arg2}
\tau_\phi = 2 - \frac{2}{d_\phi + \zeta} 
\end{equation}
where we recall that $d_\phi=d-d'$. 
Our numerical data for the point on a $d=1$ string (i.e.\ one monomer) is shown on Fig.~\ref{f:RFconjecturephi}
and we find $\tau_\phi=0.39 \pm 0.01$. If we use the best present estimate $\zeta=1.26 \pm 0.01$ 
we find $\tau_\phi^{\mathrm{conj}}=0.413 \pm 0.01$. 
If we use the value of $\zeta=1.19\pm0.01$ extracted from the scaling of $S_m$, we find $\tau_\phi^{\mathrm{conj}}=0.32  \pm 0.02$. 
While the values are roughly consistent, the precision on $\zeta$ is crucial for a precise comparison.  Inverting the conjecture (\ref{arg2}), the measurement of $\tau_\phi=0.39\pm 0.01$ leads to a conjectured $\zeta = 1.24 \pm0.01$. For a detailed discussion of
the possible artifacts we refer to  the discussion in Section \ref{a14}.

The second result is the exact expression of $\hat Z(\lambda)$ and $p_\phi(s)$ in mean field, i.e.\ for $d \geq 4$ and $d'=1$. This involves a non-trivial summation of momentum-dependent tree diagrams using instanton calculus. It yields \cite{LeDoussalWiese2008c} that
$\hat Z(\lambda)$ is given by the solution of
\begin{equation}
(\hat Z-3) \hat Z (2 \hat Z-3)=9 \lambda \ ,
\end{equation} 
which vanishes at $\lambda=0$. This yields the series expansion $\hat Z(\lambda)= \lambda + \lambda^2 
+ \frac{16}{9} \lambda^3  + \frac{35}{9} \lambda^4 + \frac{256}{27} \lambda^5 + O(\lambda^6)$. We have compared this mean-field prediction and the numerical results in $d=1$, $d_\phi=0$ on figure \ref{f:Zhat}. It is clear that loop corrections, yet to be computed, will play an important role, as was the case for bulk avalanches, see Fig.~\ref{a11}. The function 
$p_\phi(s)$, as defined here, is found to be \cite{LeDoussalWiese2008c} in mean field (i.e.\ at tree level)
\begin{equation}
   p_\phi^{\mathrm{MF}}(s) =\frac{ K_{\frac{1}{3}}\!\!\left(\frac{s}{2\sqrt{3}}\right)}{2 \pi  s}\ .
\label{217}
\end{equation}
It satisfies the normalizations (\ref{7}) and (\ref{8}), and  is related to $\hat Z (\lambda)$ via $\hat Z (\lambda) = \int_{0}^{\infty} \rmd s \, p_\phi (s)\left(\rme^{\lambda s}-1 \right)$.

Finally, the mean-field calculation \cite{LeDoussalWiese2008c} also gives:
\begin{equation}\label{aphi}
a_\phi := \frac{S_m^{\phi}}{S_m} = \frac{1}{4} 
\end{equation}
Corrections at 1-loop order slightly decrease this ratio. The $\epsilon$ expansion
predicts \cite{LeDoussalWieseToBePublished}
\begin{eqnarray}
a_\phi= \frac{1}{4} +  \alpha  \left(1 + \frac{7\pi}{6 \sqrt{3}} -\pi \right) +O(\epsilon^2) 
\end{eqnarray}
with $\alpha=-\frac{\epsilon}{3} (1-\zeta_1)$ and $\alpha = - 2/3$ here. Using the
two Pad\'e approximants gives the estimate  $1/a_\phi= 3.74 \pm 0.01$ which is consistent with the
numerically observed value of $1/a_\phi^{\mathrm{num}}=3.7 \pm 0.05$ for  $L=4000$, $m=0.00125$. 

\section{Conclusion}

In this article, we have compared the numerically obtained avalanche-size statistics at the depinning transition with the recent predictions from the functional RG based on an $\epsilon=4-d$ expansion. The critical point of the depinning  transition for  an interface of internal dimensions $d=1$ and $d=2$, driven quasi-statically in a random landscape in presence of an external quadratic well of curvature $m^2$, is reached  in the limit of $m\to 0$. We have shown that the avalanche-size distribution $P(S)$ takes the expected scaling form with the upper cutoff scale $S_m \sim m^{-d - 2 \zeta}$, involving a universal function $p(s)$ in the rescaled variable $s=S/S_m$. As we confirmed, it does not depend on whether the microscopic disorder is of random-field or random-bond type. We have computed numerically the function $p(s)$, its moments and its characteristic function and found in all cases good to excellent agreement with the predictions of the 1-loop FRG based on the extrapolation to $d=1$ and $d=2$. We have also studied, for $d=1$, the {\it local} avalanches, and there too, we found a rather satisfactory agreement with available analytical predictions. However, it remains an outstanding challenge to compute the local avalanche-size distribution within the FRG beyond mean-field. 

Some fine points deserve discussion and further study. First we have not found any clear-cut signature that the conjecture for the avalanche exponent $\tau$ be violated at depinning. However,  we can not rule out such a violation below a $0.02$ precision in $\tau$. A better numerical determination of $\tau$, comparable in precision to the one which exists for the roughness $\zeta$ at $m=0$ would be crucial to confirm or invalidate the conjectured relation between $\tau$ and $\zeta$.  Presently, the mass, i.e.\ the quadratic well, appears necessary for a proper definition of the steady state, but unfortunately, this hampers the attempts at a more precise determination of $\tau$.

Second, there has been a recent proposal, in the case of the random-field Ising model \cite{LiuDahmen2006}, that avalanche-size distributions for statics and depinning are described by the same universal functions. Although  the physics underlying this hypothesis is not clear to us, one may still ask the question for the present model \footnote{This hypothesis is proven incorrect in $d=0$ for the class of short-range correlated forces (random field class)  where one finds on one hand the Sinai-model class in the statics \cite{LeDoussal2008} and on the other, the Gumbel extremal statistics class at depinning \cite{LeDoussalWiese2008a}, with vastly different distributions in each class.}. One may for instance compare our present results to the one in the statics in \cite{LeDoussalMiddletonWiese2008}. Currently, our precision is not sufficient to conclude. For instance, in $d=2$ and for RF disorder, the conjecture (\ref{conj}) for $\tau$ gives $\tau=1.25$ for the statics and $\tau=1.2735\pm 0.0005$ for depinning, which are difficult to distinguish numerically.  We simply note that both statics and depinning data are in good agreement with extrapolations from the 1-loop FRG \cite{LeDoussalWiese2008c,FedorenkoLeDoussalWieseInPrep}, but it remains to be analyzed at two loops. As noted previously, since the roughness exponents are different, if the conjecture holds both in the statics and driven dynamics, then the avalanche-size exponents, and presumably the associated distributions, cannot be the same. We leave these subtle questions for the future.

To conclude, it is highly satisfactory that the functional-RG field theory for the avalanche statistics passes all numerical tests. Other interesting observables can now be computed numerically,  and studied on a more solid footing, such as the distribution of lateral sizes, or correlations between avalanches. These provide a motivation to further develop the theory. Finally we hope that our present work will motivate similar studies in experiments. 

\acknowledgements
This work was supported by
ANR under program 05-BLAN-0099-01.


\end{document}